\documentclass[fleqn,usenatbib,usedcolumn]{mnras}
\usepackage{graphicx,graphics,amsmath,amssymb,bm,calligra}
\usepackage{psfrag,epsfig,natbib,hyperref,tabularx,colortbl}
\usepackage{color}
\usepackage[T1]{fontenc}
\usepackage{ae,aecompl}
\usepackage[table,xcdraw]{xcolor}
\usepackage{mathptmx,pdflscape}
\usepackage{multicol,multirow}
\usepackage{txfonts}
\usepackage{mathrsfs}
\usepackage{epstopdf}
\definecolor{LightCyan}{rgb}{0.88,1,1}
\def\HI{{\ion{H}{I}~}}
\def\HII{{\ion{H}{II}~}}
\def\xb{\bar{x}_{\rm \ion{H}{I}}}
\def\xion{{x_{\rm i}}}

\def\tTb{\tilde{T}_{\rm b}}

\def\kk{\mathbfit{k}}
\def\xx{\mathbfit{x}}
\def\cov{\mathbfss{C}}
\def\covp{\mathscr{C}}

\def\signal{EoR 21-cm signal}

\def\tri{\bar{T}(k_i,k_j)}

\def\mpc{~\rm Mpc}
\def\impc{~{\rm Mpc}^{-1}}
\def\Mmin{M_{\rm min}}
\def\Nion{N_{\rm ion}}
\def\Rmfp{R_{\rm mfp}}
\def\DM{\Delta M_{\rm min}/M_{\rm min}}
\def\DN{\Delta N_{\rm ion}/N_{\rm ion}}
\def\DR{\Delta R_{\rm mfp}/R_{\rm mfp}}
\def\Nk{N_{\rm k_{i}}}
\defcitealias{shaw_2019}{Paper I}

\title[EoR parameter estimation]{The impact of non-Gaussianity on the Epoch of Reionization parameter forecast using 21-cm power spectrum measurements}

\author[A. K. Shaw, S. Bharadwaj and R. Mondal]{
Abinash Kumar Shaw,$^{1}$\thanks{E-mail:\href{mailto:abinashkumarshaw@iitkgp.ac.in}{abinashkumarshaw@iitkgp.ac.in}} Somnath Bharadwaj$^{1}$ and Rajesh Mondal$^{2}$\\
$^{1}$Department of Physics \& Centre for Theoretical Studies, Indian Institute of Technology Kharagpur, Kharagpur, India -- 721302\\
$^{2}$Astronomy Centre, Department of Physics and Astronomy, University of Sussex, Brighton BN1 9QH, UK
}

\date{Accepted XXX. Received YYY; in original form ZZZ}
\pubyear{2020}

\begin{document}
\label{firstpage}
\pagerange{\pageref{firstpage}--\pageref{lastpage}}
\maketitle

\begin{abstract}
Measurements of the Epoch of Reionization (EoR) 21-cm signal hold the potential to constrain models of reionization. In this paper we consider a reionization model with three astrophysical parameters namely (1) the minimum halo mass which can host ionizing sources, $\Mmin$, (2) the number of ionizing photons escaping into the IGM per baryon within the halo, $\Nion$ and (3) the mean free path of the ionizing photons within the IGM, $\Rmfp$. We predict the accuracy with which these parameters can be measured from future observations of the 21-cm power spectrum (PS) using the upcoming SKA-Low. Unlike several earlier works, we account for the non-Gaussianity of the inherent EoR 21-cm signal. Considering cosmic variance only and assuming that foregrounds are completely removed, we find that non-Gaussianity increases the volume of the $1 \sigma$ error ellipsoid of the parameters by a factor of $133$ relative to the Gaussian predictions, the orientation is also different. The ratio of the volume of error ellipsoids is $1.65$ and $2.67$ for observation times of $1024$ and $10000$ hours respectively, when all the $\kk$ modes within the foreground wedge are excluded. With foreground wedge excluded and for $1024$ hours, the 1D marginalized errors are $(\DM,\DN,\DR)=(6.54, 2.71, 7.75) \times 10^{-2}$ which are respectively $2 \%$, $5 \%$ and $23 \%$ larger than the respective Gaussian predictions. The impact of non-Gaussianity increases for longer observations, and it is particularly important for $\Rmfp$.
\end{abstract}


%======================================================================================%

\section{Introduction}\label{sec:intro}
The Epoch of Reionization (EoR) is an important but largely unexplored phase of cosmic history. The baryons, predominantly atomic hydrogen, go through a phase change from the neutral state (\ion{H}{I}) to an almost ionized state (\ion{H}{II}) during this epoch. Our understanding of the EoR till now depends on a few indirect observations. The intergalactic medium (IGM) is nearly completely ionized at present. However the measurements of Gunn-Peterson optical depth $\tau_{\rm GP}$ (corresponding to Ly$\alpha$), using the observed spectra of high-$z$ quasars, show a rise in the value of $\tau_{\rm GP}$ with $z$ (e.g. \citealt{becker_2001,Fan_2002,Fan_2006,gallerani_2006,Becker_2015}). The complete Gunn-Peterson troughs seen in  $z \gtrsim 6$ quasar spectra indicate the IGM to be sufficiently neutral ($\xb\simeq 10^{-5}-10^{-4}$) at these epochs. These observations suggest that the reionization would have been completed by around $z = 6$ \citep{McGreer_2014}. Also the free electrons in the ionized IGM interact with the Cosmic Microwave Background (CMB) photons through Thomson scattering, and CMB experiments measure the corresponding optical depth $\tau_{\rm Th}$. One can estimate the redshift corresponding to the beginning of the EoR based on an appropriate reionization model. Considering different models of reionization, the latest measurement of $\tau_{\rm Th}=0.054\pm 0.007$ suggests that the IGM would have been $10\%$ ionized at $z\sim 10$ \citep{Planck_Cosmo_2018}.

The recent studies on the Ly$\alpha$ emitters (LAEs) at high redshift (e.g. \citealt{Malhotra_2004,Hu_2010,Kashikawa_2011,Jensen_2013}) provide another probe to study the reionization indirectly. A decrease in the Ly$\alpha$ luminosity function (LF) has been observed while moving from $z=6$ to $z=8$ whereas the Ly$\alpha$ clustering does not evolve much in the same redshift range \citep{Jensen_2014,santos_Lya-2016}. This implies that the IGM was considerably neutral ($\xb=0.2$) and patchy at $z\ge 7$ and it becomes mostly ionized ($60-80\%$) at $z\sim 7$ \citep{Ouchi_2010,Faisst_2014,Konno_2014,Ota_2017,Zheng_2017}. The study of the damping wings present in the high redshift ($z\gtrsim 7)$ quasar spectra (e.g. \citealt{Greig_2016,Greig_2019a,Davies_2018,Wang_2020,Durovcikova_2020,Reiman_2020}) also provides similar results regarding the neutral state of the IGM. A recent study of UV-LF of the `oligarchs' \citep{Naidu_2020} has measured IGM neutral fraction to be $(0.9,0.5,0.1)$ at $z= (8.2,6.8,6.2)\pm0.2$ that suggests a rapid reionization. Moreover the study of the UV-LF of the Lyman Break Galaxies at high redshift ($z\gtrsim7$) also support a similar rapid reionization scenario (e.g. \citealt{Mason_2018,Mason_2019,Hoag_2019}). All these indirect experiments commonly suggest that the reionization continues in range $6\leq z \leq 12$ (e.g. \citealt{Robertson_2013,Robertson_2015,Mondal_I,Mitra_2017,Mitra_2018,Dai_2019}). However these indirect observations loosely constrain the EoR and are unable to provide a strong insight to the physics behind the reionization such as the generic characteristics of the ionization sources, the accurate timing and the span of the EoR and the topology of the $\HI$ brightness temperature maps, etc.

After the recombination epoch, the CMB hardly interacts with the neutral intervening medium. This restricts the CMB from probing the evolution of the structures till the end of EoR. The 21-cm radiation, which is involved in the hyperfine transition of \HI, is a promising probe to study the high redshift universe including EoR (e.g. \citealt{sunyaev_1972,Hogan_1979}). There are existing and the upcoming radio interferometers aiming to observe the brightness temperature fluctuations of the redshifted 21-cm signal from EoR which we coin as the `EoR 21-cm signal'. However the detection of the signal is not yet possible due to the foreground contamination from galactic and extra-galactic source. The foregrounds are $\sim 10^4-10^5$ times stronger (e.g. \citealt{Ali_2008,Bernardi_2009,Bernardi_2010,Abhik_2012,Paciga_2013,Beardsley_2016}) compared to the signal. The foregrounds, system noise and calibration errors together keep the current observations at bay from directly detecting the EoR 21-cm signal. As a consequence, the first detection is likely to be statistical in nature.  These observations plan to measure the power spectrum (PS) of the \signal~(e.g. \citealt{Bharadwaj2001,Ali_2004,Bharadwaj_2005}). Several radio interferometers such as the GMRT\footnote{\url{http://www.gmrt.ncra.tifr.res.in}} \citep{Swarup_1991}, LOFAR\footnote{\url{http://www.lofar.org}} \citep{vanHaarlem_2013},  MWA\footnote{\url{http://www.haystack.mit.edu/ast/arrays/mwa}} \citep{Tingay_2013} and PAPER\footnote{\url{http://eor.berkeley.edu}} \citep{Parsons_2010} have carried out observations to measure the EoR 21-cm PS.  However, only few weak upper limits on the PS amplitudes have been reported in the literature to date (e.g. GMRT: \citealt{Paciga_2011}, \citealt{Paciga_2013}; LOFAR: \citealt{Yatawatta_2013}, \citealt{Patil_2017}, \citealt{Gehlot_2019}, \citealt{LOFAR_Mertens_2020};  MWA: \citealt{Dillon_2014}, \citealt{Jacobs_2016}, \citealt{Li_2019}, \citealt{Barry_2019}, \citealt{Trott2020}; PAPER: \citealt{Cheng_2018}, \citealt{Kolopanis_2019}). A few more upcoming telescopes with improved sensitivity such as HERA\footnote{\url{http://reionization.org}} \citep{DeBoer_2017} and  SKA\footnote{\label{tel:ska}\url{http://www.skatelescope.org}} \citep{AASKA14} also aim to measure the EoR 21-cm PS. Apart from PS, several other estimators such as the variance \citep{patil_2014}, bispectrum \citep{Bharadwaj_Pandey,Yoshiura_2015,bispec_shimabukuro,Majumdar_2018} and Minkowski functional \citep{Akanksha_2017,Bag_2018,Bag2019,Kapahtia_2019} are being used to quantify the \signal. These estimators are supposed to be rich in information about the underlying physical processes during EoR.

There could be several physically motivated processes which drive the ionization of \HI in the universe and a few known processes have already been modelled through parameters. These parameters, which may affect the measured estimators (here the 21-cm  PS), are typically related to the generic properties of the first ionizing sources and the state of IGM during reionization. A precise study of these parameters is mandatory to build a deep insight to the EoR. The main issue is related to the question ``How well can one constrain the reionization physics through model parameters given direct EoR observations?''. Several previous studies (e.g. \citealt{21cmmc,Ewall-Wice,Binnie_2019,Greig_2019b, Park_2019}) have tried to put constraints over various reionization models for different ongoing and upcoming radio experiments which are devoted for the EoR observations. Since we are taking help of the statistical estimator (mainly PS) of the EoR 21-cm signal, the uncertainties in the measured 21-cm PS will translate into the uncertainties in the inferred parameters.

The parameter estimation using observables are conventionally done using Bayesian statistics \citep{Sanjib_2017} in two separate ways in cosmology. (1) The Fisher matrix formalism, which provides a general theory to compute the probability distribution of the parameters given an observed data. This formalism is a powerful tool to interpret the observed data, however it fails when a simple analytic solution does not exist. (2) The Markov Chain Monte Carlo (MCMC) technique, which is a brute-force technique that samples parameters from a specific distribution for a given observed data set. There are several works that have employed Fisher formalism (e.g. \citealt{McQuinn_2006,Mao_2008, Pober_2014,Ewall-Wice,bispec_shimabukuro,Binnie_2019}) and several others that have used the MCMC (e.g. \citealt{patil_2014,21cmmc,Hassan_2017,Kern_2017,Cohen_2018,Greig_2018,Greig_2019b,Park_2019}) to study the sources and physical processes responsible for reionization. Recently, the use of machine learning has become popular in cosmology and there are few works which have tried to study reionization with the help of artificial neural networks (e.g. \citealt{Schmit_2017,Shimabukuro_2017,Hassan_2018,Doussot_2019,Gillet_2019,Florian_2020}). Even though predictions by the neural networks are fast enough its training is still computationally expensive and time consuming. Besides any bias in the training set data may change the results. We choose to employ Bayesian Fisher matrix formalism for the purpose of our analysis.

Recent simulations of the EoR 21-cm signal \citep{Mondal_2015,Mondal_I} show that the signal is inherently non-Gaussian. The non-Gaussianity introduces a non zero trispectrum contribution to the error variance of the measured 21-cm PS. The authors in \citet{Mondal_II} have explicitly shown that the non-Gaussianity raises the cosmic variance (CV) of the 21-cm PS a few thousand times relative to the Gaussian estimates of CV at large $k$ modes and towards the end of the reionization ($z\simeq7$). \citet{shaw_2019}, hereafter denoted as \citetalias{shaw_2019}, have recently investigated the effects of non-Gaussianity on the total error covariance (including system noise and foregrounds) of the 21-cm PS during an observation. They find that the impact of non-Gaussianity in total error variance is relatively less prominent once observations are considered. However, trispectrum contribution is found to be important in range $k\simeq 0.1-1\impc$ during later stages ($z\leq 8$) of reionization. The aforementioned works on the predictions of constraining the EoR using measurements of the 21-cm PS have frequently treated the EoR 21-cm signal as a Gaussian random field. The aim of this work is to figure out the impact of non-Gaussianity over the constraints on the reionization parameters.

We study the effects of non-Gaussianity on reionization parameter estimation in the context of a future radio observation using SKA-Low \citep{SKA_Low_v2}. This experiment is planned in Australia with a  station layout which consists of a compact core and three spiral arms. The arms will have extent that can provide antenna separations up to $\sim 64$ km. This interferometer is an array of $512$ stations, each of which is a collection of several log-periodic dipole antennas having both the polarizations and placed within a circle of diameter $\sim35~{\rm m}$. It will have a considerably large field of view (FoV) $\sim 20~{\rm deg}^2$ on the sky. Owing to its large frequency bandwidth in range $50-350~{\rm MHz}$, SKA-Low will be able to probe 21-cm signal within a redshift range $3\leq z\leq 27$ that includes the Cosmic Dawn\,(CD), EoR and a part of post-reionization epoch. This is going to be the most sensitive radio interferometer to date. In our analysis, we consider deep observation of a particular field to achieve greater sensitivity \citep{Greig_2019b}. 

This paper discusses the prospects of measuring the reionization model parameters using the upcoming SKA-Low observations and also elaborates about the impact of non-Gaussianity of the 21-cm signal. Here we employ the Fisher matrix formalism to achieve our goal of constraining reionization through the model parameters. The structure of this paper is as follows. Section~\ref{sec:sim} provides a brief discussion on the reionization simulation and its model parameters. Next, a detailed description of our methodology is presented in Section~\ref{sec:method}. The findings from our analysis is shown in Section~\ref{sec:res} followed by the summary and discussion in Section~\ref{sec:summ}. Our simulation uses the best fitted cosmological parameters from Planck+WP observations \citep{planck_2014}.

%======================================================================================%

\section{Simulating the redshifted 21-\lowercase{cm} signal from E\lowercase{o}R}\label{sec:sim}
In this paper, we use an ensemble of \signal~ simulated at the six different redshifts $z=13,~11,~10,~9,~8$ and $7$. The simulation employs a semi-numerical technique \citep{Majumdar_2013,Mondal_2015} to generate the redshifted 21-cm brightness temperature fluctuations. The simulation procedure can be divided into three major steps. The first step is to simulate the dark matter density field using a particle mesh \textit{N}-body code \citep{Srikant2004}. The dark matter density field is generated within a comoving box of volume $V=[215.04\mpc]^3$ with the spatial resolution of $0.07\mpc$ and the mass resolution $1.09\times 10^8~{\rm M}_\odot$. In the second step, we identify the dark matter halos using the Friends-of-Friend (FoF) algorithm with a linking length of $0.2$ times the mean inter-particle separation. We only consider  halos consisting of a minimum of $10$ dark matter particles which corresponds to minimum  halo mass of $1.09\times 10^9~{\rm M}_\odot$ in our simulations. The third step in our simulations is to generate the \HI 21-cm brightness temperature map using a reionization model which closely follows that in \citet{tirh_2009}. The \textit{N}-body, FoF and reionization codes are all publicly available\footnote{\url{https://github.com/rajeshmondal18/}}. 

The reionization model used here has two basic assumptions. The first assumption is 
that the hydrogen gas follows the underlying dark matter distribution,  and the second assumption is that the sources of ionizing ultraviolet (UV) radiation are located within the dark matter halos. Here we consider UV photo-ionization of the hydrogen in the IGM to be the only process that  drives reionization. We model the reionization process using three physical parameters which are the minimum halo mass $\Mmin$, the ionization efficiency $\Nion$ and the mean free path of ionizing photons $\Rmfp$. We provide detailed descriptions of these parameters in subsequent paragraphs. 

\begin{itemize}
\item {$\bm{\Mmin:}$} This is the lowest halo mass above which a halo can accrete sufficient hydrogen for sustained star formation. The first stars form in metal free environments which  requires hydrogen to cool either  via atomic cooling or through molecular line cooling  in highly dense clumps (e.g. \citealt{Yoshida_2012,Klessen_2018}). On the other hand, the UV photons from the stars photo-evaporate the hydrogen gas from the clumps as soon as they form. Previous studies show that the halos having a virial temperature $T_{\rm vir}\geq 10^4~{\rm K}$ are able to sustain the cooling of hydrogen clumps against the photo-evaporation process. Observations \citep{Bolton_2007} suggest that EoR is ``photon-starved'' and extended  ending  at $z\sim6$. This implies that a sufficient number of ionizing photons are required from halos of various masses in order to complete the reionization process by $z\sim6$. Decreasing the value of $\Mmin$ while keeping the other parameters fixed would result in more number of ionizing photons from the smaller halos. This causes EoR to end before $z\sim 6$ whereas  increasing $\Mmin$ delays the reionization process. The reionization simulations of \citet{tirth_2008} suggest that $\Mmin \sim 10^6-10^7~{\rm M}_\odot$ is required to produce the Thomson scattering optical depth of IGM and the Gunn-Peterson troughs consistent with observations. However their simulations do not include the metal-free Population III stars which are highly efficient sources of reionizing photons. The value of $\Mmin$ is expected to increase if Population III stars are also included. However,  more recent simulations by \citet{Finlator_2016} has constrained the value of $\Mmin \sim 10^9~{\rm M}_\odot$ using the observed UV luminosity function  in the redshift range $6\leq z \leq 8$ (well within the EoR). We have chosen a fiducial value of $\Mmin=1.09\times 10^9~{\rm M}_\odot$ for our simulations \citep{Mondal_II}.

\item {$\bm{\Nion:}$} Our model assumes that total number of ionizing UV photons which escape into the IGM from a halo of mass $M_h$ is directly proportional to $M_h$. The proportionality relation can be expressed as (eq. 3 of \citealt{Majumdar_2014})
\begin{equation}
N_\gamma(M_h)=\Nion \frac{M_h}{m_{\rm p}} \frac{\Omega_b}{\Omega_m}~,
\label{eq:nion}
\end{equation}
where $\Nion$ is a dimensionless proportionality constant which quantifies the number of ionizing photons escaping into the IGM per baryon within the halo. This parameter primarily depends upon the properties of the ionizing sources and several other degenerate factors such as the star formation efficiency $f_*$, escape fraction of ionizing photons from a halo $f_{\rm esc}$,  and the hydrogen recombination rate \citep{RoyChoudhuri_2009}. Studies  show that the value of $\Nion$ is expected to evolve with redshift, however we do not expect this to drastically modify the reionization scenario (e.g. \citealt{Naidu_2020}). We have used a fiducial value $\Nion=23.21$ throughout this work. This provides a scenario where the reionization of the IGM   starts at $z\sim 13$, becomes $50\%$  at $z\simeq 8$ and ends by $z\sim 6$. An increment in the value of $\Nion$ will hasten the process of reionization and vice-versa.

\item {$\bm{\Rmfp:}$} The mean free path of the ionizing photons is the third physical parameter which governs the typical size of \HII regions, particularly before they overlap. $\Rmfp$ typically depends upon the density and the distribution of the Lyman limit systems in the IGM. The observations of such systems suggest $\Rmfp$ will have values in the range $3-80\mpc$ at $z\sim 6$ \citep{cowie}. However, recent simulations of \citet{Sobacchi_2014} show that inhomogeneous recombination limits the values within a smaller range $5-20\mpc$ and we have  chosen  a fiducial value $\Rmfp=20\mpc$ which is in agreement with this. 
\end{itemize}

Our semi-numerical reionization code is based on the excursion set formalism of \citet{Furlanetto_2004}. Considering a grid point $\xx$, the number  density of the ionizing photons $\langle n_\gamma(\xx) \rangle_R$ smoothed over a sphere of comoving radius $R$ is compared with the corresponding smoothed number density of hydrogen $\langle n_{\rm H}(\xx) \rangle_R$. The comparison is done  varying the radius $R$ in steps starting from a minimum value which is the grid size to a maximum value of the photon mean free path $\Rmfp$. The grid point is said to be completely ionized if it satisfies the condition (eq. 4 of \citealt{Majumdar_2014})
\begin{equation}
\langle n_\gamma(\xx) \rangle_R \geq \langle n_{\rm H}(\xx) \rangle_R\,,
\label{eq:condition}
\end{equation}
at any step, and the corresponding ionized fraction is set to $\xion=1$. If the above condition remains unsatisfied for $R\leq \Rmfp$, the grid is partially ionized and assigned  a value $\xion=\langle n_\gamma(\xx) \rangle_R/\langle n_{\rm H}(\xx) \rangle_R$ where the smoothing radius $R$ is equal to the grid size.

We have followed the methodology of \citet{Majumdar_2013} to apply redshift space distortion  to the resulting \HI map,  and the final 21-cm brightness temperature map is produced on a grid that is eight times coarser as compared to that of the \textit{N}-body simulation. We have generated an ensemble consisting $50$ statistically independent realizations of the EoR 21-cm signal, all corresponding to the fiducial values of the parameters $[\Mmin,~\Nion,~\Rmfp]=[1.09 \times 10^9 \, {\rm M}_{\odot},~23.21,~20 \,\mpc] $. This ensemble was used to evaluate the 21-cm power spectrum and trispectrum pertaining to the fiducial model. Note that this ensemble is the same as that which has been used in \citet{Mondal_II} and \citet{shaw_2019}. The integrated Thomson scattering optical depth computed for our fiducial model is $\tau=0.057$ which is consistent with the observations \citep{Planck_Cosmo_2018} where $\tau=0.054\pm 0.007$.

We quantify the statistics of the \signal~ using its power spectrum (PS) which is the primary observable of reionization experiments. The EoR 21-cm PS at a particular wave number $\kk$ is $P(\kk)=V^{-1} \langle \tTb(\kk) \tTb(-\kk) \rangle$ where $V$ is the simulation (or observation) volume, $\tTb(\kk)$ is the Fourier transform of the 21-cm brightness temperature fluctuations and $\langle \cdots \rangle$ denotes the ensemble average. We use the bin-averaged EoR 21-cm PS (averaged within semi-spherical bins in $k$ space) which, for an $i$-th bin, is given as (see eqs. 20 and 22 of \citealt{Mondal_I})
\begin{equation}
\bar{P}(k_i) =  \frac{1}{\Nk} \sum_{a \in i}^{} P(\kk_a)~.
\label{eq:bin-avg}
\end{equation}
Here  the sum $\sum_a$ is over all the $\kk_a$ modes within the $i$-th bin, $\Nk$  is the number of modes in the bin and $k_i$ the average comoving wave number corresponding to the bin. 

It is necessary to consider higher order statistics in order to quantify the effects of non-Gaussianity on the EoR 21-cm signal PS error covariance. This non-Gaussianity manifests itself as a non-zero trispectrum $T(\mathbfit{a},\mathbfit{b},\mathbfit{c},\mathbfit{d})$ which, using  $\mathbfit{a}$ to denote $\kk_a$, is defined through
\begin{equation}
\begin{split}
\langle \tTb(\mathbfit{a})\tTb(\mathbfit{b}) \tTb(\mathbfit{c}) \tTb(\mathbfit{d}) \rangle & =V  \,  \delta_{\mathbfit{a}+\mathbfit{b}+\mathbfit{c}+\mathbfit{d},0} \,  T(\mathbfit{a},\mathbfit{b},\mathbfit{c},\mathbfit{d}) \\
&+ V^2 \times [\delta_{\mathbfit{a}+\mathbfit{b},0} \delta_{\mathbfit{c}+\mathbfit{d},0}P(\mathbfit{a})P(\mathbfit{c}) \\
&+\delta_{\mathbfit{a}+\mathbfit{c},0} \delta_{\mathbfit{b}+\mathbfit{d},0}P(\mathbfit{a})P(\mathbfit{b})\\
&+\delta_{\mathbfit{a}+\mathbfit{d},0} \delta_{\mathbfit{b}+\mathbfit{c},0}P(\mathbfit{a})P(\mathbfit{b})] ~,
\end{split}
\label{eq:tri}
\end{equation}
In our analysis we have used the bin-averaged trispectrum. Considering a pair of bins namely $i$ and $j$, this is defined as 
\begin{equation}
\tri= \frac{1}{\Nk N_{\rm k_{j}}} \sum_{a\in i , b \in j}^{} T(\mathbfit{a},-\mathbfit{a},\mathbfit{b},-\mathbfit{b})~,
\label{eq:bin_trispec}
\end{equation}
where the two wave vectors $\mathbfit{a}$ and $\mathbfit{b}$ lie within the $i$-th and the $j$-th bins respectively. \citet{Mondal_I} have used the ensemble described earlier to indirectly estimate the bin averaged trispectrum, and we have used this for our work here. 

\begin{figure}
    \begin{center}
    \includegraphics[scale=0.58,angle=0]{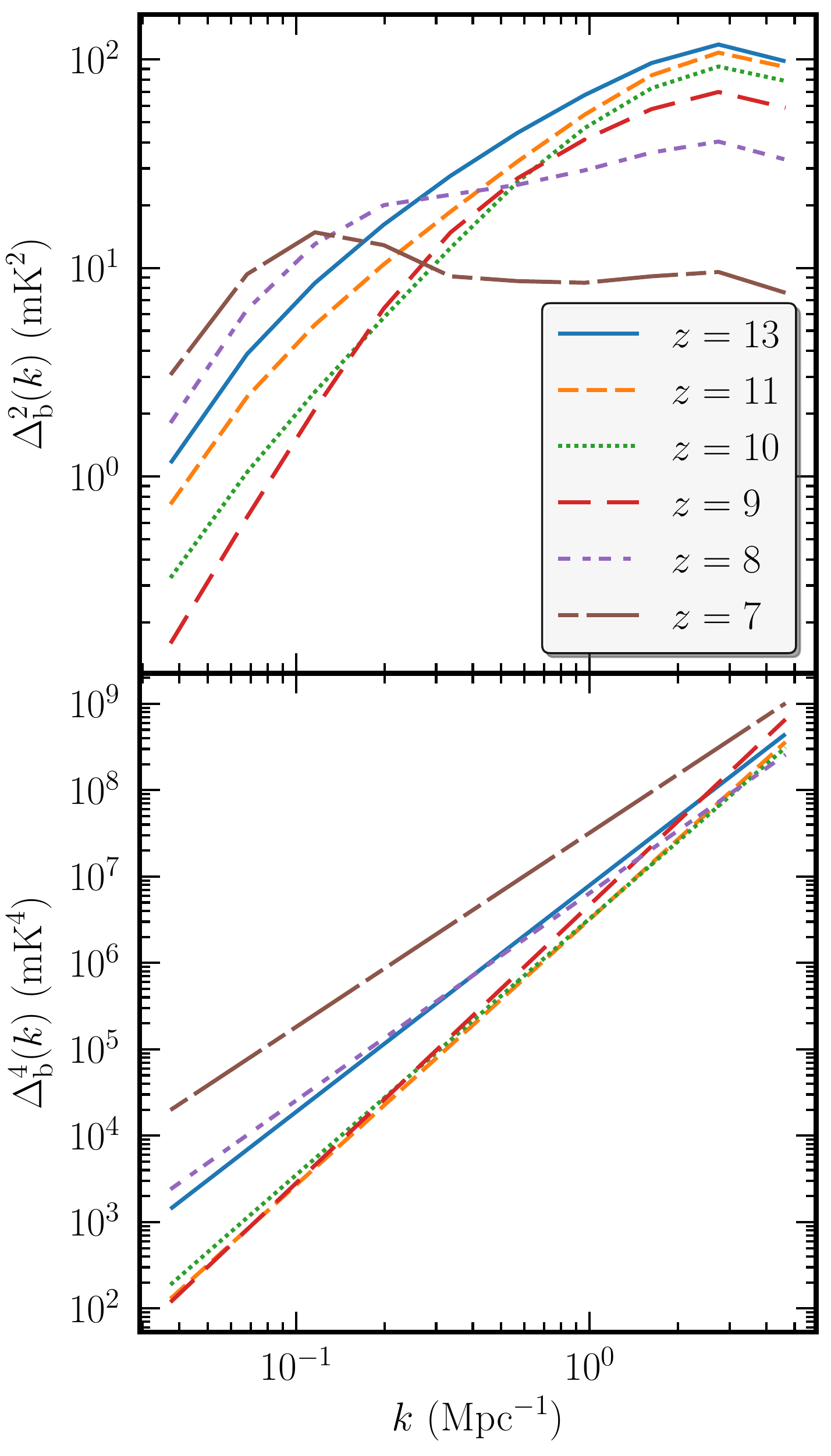}
    \caption{The dimensionless bin-averaged \HI 21-cm PS $\Delta_b^2(k)$ (top panel) and the corresponding dimensionless bin-averaged trispectrum $\Delta_b^4(k)$ (bottom panel). Different lines correspond to the comoving slices at six different redshifts.}
    \label{fig:pk_tri}
    \end{center}
\end{figure}

The upper panel of Figure \ref{fig:pk_tri}  shows   the dimensionless bin-averaged 21-cm PS $\Delta^2_b(k)=k^3\bar{P}(k)/(2\pi^2)$ as a function of $k$ at the six redshifts which we have considered for our analysis. Several of the features visible in the 21-cm  PS are sensitive to the values of the model parameters.  We have quantified  this dependence  in subsequent parts of this paper.

The lower panel of Figure \ref{fig:pk_tri} shows the diagonal elements of the dimensionless bin-averaged trispectrum $\Delta^4_b(k)=k^9\bar{T}(k,k)/(2\pi^2)$ as a function of $k$ at the six redshifts 
which we have considered for our analysis.   The power spectrum and the trispectrum shown here have both been used to  calculate the error covariance  matrix for measuring the power spectrum. 

%======================================================================================%
\section{Methodology}\label{sec:method}
Radio interferometric observations will result in a measurement of the bin-averaged EoR 21-cm PS $\bar{P}(k_i)$. The errors expected in this measurement can be quantified through the error covariance matrix 
\begin{equation}
\cov_{ij} =\langle [\Delta \bar{P}(k_i)] [\Delta \bar{P}(k_j)] \rangle \,,
\label{eq:cov0}
\end{equation}
where $\Delta \bar{P}(k_i)=\hat{P}(k_i)-\bar{P}(k_i)$ and $\hat{P}(k_i)$ is the binned PS estimator. The diagonal element $\cov_{ii}$ quantifies the error variance of the 21-cm PS measured at the $i$-th bin, whereas the off-diagonal terms ($i \neq j$) quantify the correlations between the errors in the different $k$ bins. Here we consider the situation where the measured 21-cm PS is used to constrain the values of the parameters $q_\alpha=[\Mmin,\,\Nion,\,\Rmfp]$ of our reionization model. This can be achieved by finding the parameter values for which the model predictions best match the measured 21-cm PS. The errors in the measured 21-cm PS will be reflected in the error estimates for the best fit parameter values. Here we employ the Fisher matrix formalism to predict the errors expected in the estimated parameter values. We note that the usage of the Fisher matrix assumes Gaussian errors on the estimated model parameters.
The Fisher matrix $ F_{\alpha \beta}$ corresponding to the model parameters $q_{\alpha}$ is related to the error covariance $\cov_{ij}$ through (e.g. \citealt{Repp_2015})
\begin{equation}
    F_{\alpha \beta}= \sum_{i,j} \left( \frac{\partial \bar{P}(k_i)}{\partial q_\alpha} [\cov^{-1}]_{ij} \frac{\partial \bar{P}(k_j)}{\partial q_\beta}\right)~,
    \label{eq:fisher}
\end{equation}
where the summation is over all the $k$ bins at which the 21-cm PS is measured, and ${\partial \bar{P}(k_i)}/{\partial q_\alpha}$ is the derivative of the model prediction with respect to the model parameters. The derivative here quantifies how sensitive the different features seen in the bin-averaged 21-cm PS (Figure \ref{fig:pk_tri}) are with respect to changes in the various model parameters. The Cram\'er-Rao inequality \citep{rao_1945,cramer,Kay} implies that the inverse of the Fisher matrix $F_{\alpha \beta}$ provides an estimate of the lower bound of the  error covariance  $\covp_{\alpha\beta}$ of the parameters $q_{\alpha}$. In the present work we have used 
\begin{equation}
    \covp_{\alpha\beta}= [F^{-1}]_{\alpha \beta}~,
    \label{eq:cr}
\end{equation}
whereby the results presented here may be interpreted as lower bounds for the error covariance $\covp_{\alpha\beta}$. We have used equations (\ref{eq:fisher}) and (\ref{eq:cr}) to estimate the lower bounds on the errors for measuring our model parameters using future observations with SKA-Low.

\subsection{Computing the EoR 21-cm PS error covariance}\label{sub:cov}
The error covariance of the 21-cm PS arises from  two distinct contributions namely the cosmic variance (CV) which is the statistical uncertainty inherent to the signal, and the system noise which arises due to the instrument and the sky temperature.  We briefly discuss the contributions from these two components separately.

\subsubsection{Cosmic variance}
The cosmic variance (CV) quantifies the uncertainty  which is inherent to the signal. We can write the CV of the bin-averaged 21-cm PS $\bar{P}(k_i)$ as \citep{Mondal_I,shaw_2019}
\begin{equation}
\cov_{ij} = \frac{[\bar{P}(k_i)]^2}{\Nk} \delta_{ij} + \frac{\tri}{V} \,.
\label{eq:cov_CV} 
\end{equation}
The analysis can be considerably simplified in the situation where the 21-cm signal is assumed to be a Gaussian random field for which the trispectrum vanishes $(\tri=0)$, and we have  
\begin{equation}
\cov_{ij}^{\rm G}= \frac{[\bar{P}(k_i)]^2}{\Nk} \delta_{ij}~.
\label{eq:cov-gauss}
\end{equation}
As obvious from the equation (\ref{eq:cov-gauss}), the error covariance matrix for a Gaussian random field is diagonal \textit{i.e.} the errors in the PS measured in different $k$ bins are uncorrelated. Each diagonal element $\cov_{ii}$ quantifies the variance of the error in the measured 21-cm PS in the respective bin. The error variance is proportional to square of the estimated 21-cm PS $[\bar{P}(k_i)]^2$ and inversely proportional to ${\Nk}$ the number of independent $\kk$ modes in the bin. 

The EoR 21-cm signal is a highly non-Gaussian field \citep{Bharadwaj_Pandey}, and it is necessary to consider the higher order statistics. \citet{Mondal_2015} have shown that the non-Gaussianity considerably affects the PS error estimates (i.e. cosmic variance), and it is not possible to achieve an SNR above a certain limiting value, even by increasing the number of Fourier modes in a $k$ bin. Their analytical model also shows that the error variance gets additional contribution from the non-zero trispectrum which leads to larger error variance as compared to the Gaussian predictions. The trispectrum also introduces non-zero off-diagonal terms. \citet{Mondal_I,Mondal_II} have found statistically significant ($>1\sigma$) correlations and anticorrelations which depend on the considered length-scales and also the stage of reionization. The trispectrum arising from the non-Gaussianity of the EoR 21-cm signal has a substantial effect on the PS error covariance matrix (equation~\ref{eq:cov_CV}), and the earlier works mentioned above have studied this in considerable detail.

\subsubsection{System noise}
We have considered the proposed SKA-Low antenna layout \citep{SKA_Low_v2} for which we have simulated the distribution of antenna pair separations  $\mathbfit{d}$ (see e.g. Figure~8 of \citealt{Mondal_2019b}) corresponding to $8$ hours of observations with an integration time of $60\,{\rm seconds}$ towards a fixed sky direction located at DEC$=-30^\circ$. The observations are assumed to span $N_t$ nights resulting in a total $t_{\rm obs}=N_t \times 8 \, {\rm hours}$ of observations. In order to avoid the light-cone effect (e.g. \citealt{Light_cone_I,Light_Cone_II,Rajesh_Light-cone,Mondal_2019a}), the subsequent analysis is restricted to slices of width $\Delta z=0.75$ centered at each of the six redshifts mentioned earlier. Each slice has the visibility measurements at the simulated baselines $\mathbfit{U}=\mathbfit{d}/\lambda_c$ where $\mathbfit{d}$ is the antenna pair separation projected on the plane perpendicular to the LoS, and $\lambda_c$ is  the wavelength that corresponds to the central frequency $\nu_c$ of the slice. Note that we restrict our analysis to the baselines within  $\mid \mathbfit{d}\mid \leq 19~{\rm km}$ as  the baseline density falls off rapidly beyond this. The  observed visibilities will provide us with measurements of the brightness temperature fluctuation  $\tilde{T}_{\rm b}(\kk)$ at $\kk=(\kk_\perp, k_\parallel)$ where $\kk_\perp = (2\pi \mathbfit{U})/(r_c \lambda_c)$ and $k_\parallel=(2 \pi m)/(r^\prime_c B)$ . Here, $B$ is the frequency bandwidth corresponding to the slice thickness $\Delta z$, $0\leq m\leq N_c/2$, $N_c=B/(\Delta \nu_c)$,  $r_c$ is the comoving distance to the centre of a slice and $r^\prime_c=(\partial r_c/\partial \nu)\mid_{\nu=\nu_c}$.

We have identified the volume of $\kk$ space corresponding to each slice and introduced a grid spanning this volume. The grid spacing on the plane perpendicular to the LoS is chosen to be $\Delta k_\perp = (2\pi D)/(r_c \lambda_c)$ whereas $\Delta k_\parallel=(2 \pi)/(r^\prime_c B)$. The visibilities measured at two different baselines at a separation $\Delta U < (2\pi D)/(r_c \lambda_c)$ are expected to be correlated \citep{Bharadwaj_2005}. The values of $\Delta k_\perp$ and $\Delta k_\parallel$ have been chosen so that each grid point has independent information. The measured visibilities are collapsed onto this grid to obtain the brightness temperature fluctuations $\tilde{T}_{\rm b}(\kk_g)$ at any grid point $\kk_g$ used for PS estimation.

In addition to the 21-cm brightness temperature fluctuations $\tTb(\kk_g)$, the total observed brightness temperature fluctuations $\tilde{T}_{\rm t}(\kk_g)$ at any grid point $\kk_g$ also has a random Gaussian system noise contribution $\tilde{T}_{\rm N}(\kk_g)$ \textit{i.e.} $\tilde{T}_{\rm t}(\kk_g)=\tilde{T}_{\rm b}(\kk_g)+\tilde{T}_{\rm N}(\kk_g)$. The corresponding noise PS is given by \citep{SumanCh_2019,shaw_2019,Mondal_2019b}
\begin{equation}
P_{\rm N}(\kk_g) = \frac{8 ~{\rm hours}}{t_{\rm obs}}\times \frac{P_0}{\tau(\kk_g)}~,
\label{eq:Pn}
\end{equation}
where $P_0$ is the system noise power spectrum for a single visibility measurement with $60 \, {\rm seconds}$ integration time. The value of $P_0$ depends on the SKA-Low antenna parameters \citep{SKA_Low_v2} and the observing frequency $\nu_c$ (see equations 1 and 2 of \citetalias{shaw_2019}). $P_0$ has values $(3.296,~2.091,~0.931,$ $0.569,~0.319~{\rm and}~0.217)\times 10^2~{\rm K}^2$ respectively at the six redshifts $(13,~11,~10,~9,~8~{\rm and}~7)$ considered here. Here we assume that it is possible to track the target field for $8$ hours each night. The resulting baseline distribution results in a non-uniform sampling of the $\kk$ space. We use $\tau(\kk_{g})$ to quantify the number of independent visibility measurements lying within a voxel centred at the grid point $\kk_g$. The system noise contribution at the different measured visibilities are uncorrelated, and consequently the noise PS falls as $1/\tau(\kk_g)$. We have used the simulated baseline distribution mentioned earlier to estimate   $\tau(\kk_g)$. The simulations used here are the same as those used in \citetalias{shaw_2019}, and the reader is referred there for further details. 

It is possible to avoid $P_{\rm N}(\kk_g)$ contribution in the estimated 21-cm PS \citep{Begum_2006,TGE}. However, it is not possible to remove the system noise contribution from the error covariance of the estimated 21-cm PS. The system noise contribution $P_{\rm N}(\kk_g)$ varies from grid point to grid point due to the non-uniform sampling. It is desirable to account for this by assigning different weights $\tilde{w}_g$ to the individual grid points $\kk_g$ when binning the 21-cm PS estimated at the different grid points. Note that introducing the weights  only affects the error covariance of the  estimated bin-averaged PS. We have chosen the weights $\tilde{w}_g$ so as to optimise the signal-to-noise ratio (SNR) for the 21-cm PS estimated in each bin. The exact analytic expression for the error covariance of the bin-averaged 21-cm PS (equation 4 of \citetalias{shaw_2019}) requires us to know the the trispectrum $T(\kk_{g_{a}} ,-\kk_{g_{a}} ,\kk_{g_{b}},-\kk_{g_{b}})$ for every pair of grid points $(\kk_{g_{a}} ,\kk_{g_{b}})$. This is  an enormous volume of information $(\sim 10^{12})$ which is beyond our scope. In \citetalias{shaw_2019} we have overcome this issue by approximating $T(\kk_{g_{a}} ,-\kk_{g_{a}} ,\kk_{g_{b}},-\kk_{g_{b}})$ using the bin-averaged trispectrum $\tri$ from \citet{Mondal_II}. In the present work we adopt Case I of \citetalias{shaw_2019} which assumes that $T(\kk_{g_{a}} ,-\kk_{g_{a}} ,\kk_{g_{b}},-\kk_{g_{b}})=\tri$ where $\kk_{g_{a}}$ and $\kk_{g_{b}}$ lie in the $i$-th and the $j$-th bin respectively. Adopting the results from \citetalias{shaw_2019}, the PS error covariance $\cov_{ij}$ for Case I is
\begin{equation}
    \cov_{ij}=\frac{1}{\sum_{g_{i}}\tilde{w}_{g_{i}}} \delta_{ij}+ \frac{\tri}{V}~,
    \label{eq:cov_obs}
\end{equation}
where $V$ is the observational volume corresponding to the telescope's FoV and Bandwidth, and  $\tilde{w}_{g_{i}}$ is the unnormalized weight at a grid point $\kk_g$ in the $i$-th bin. We obtain the weights to be
\begin{equation}
    \tilde{w}_{g_{i}}=\frac{1}{[\bar{P}(k_i)+P_{\rm N}(\kk_{g_{i}})]^2}~,
    \label{eq:wt}
\end{equation}
for which the SNR of the estimated bin-averaged PS is maximum in each  bin. Equation (\ref{eq:wt}) implies that the grid points which have more noise will contribute less to the bin-averaged PS estimation and vice-versa. Also the noise PS $P_{\rm N}(\kk_g)=\infty$ for an unsampled grid point ($\tau(\kk_g)=0$) and the associated weight becomes zero. In \citetalias{shaw_2019} we had also considered an alternative model for $T(\kk_{g_{a}} ,-\kk_{g_{a}} ,\kk_{g_{b}},-\kk_{g_{b}})$  (referred to as Case II) where we have the minimum possible correlation between the signal at different $\kk$ modes in the same bin. The readers are referred to \citetalias{shaw_2019} for a detailed discussion and a comparison of the two cases, however we have not considered Case II here.We finally note that the trispectrum  vanishes if the signal is a Gaussian random field, and in this  situation the results are the same for both Case I and II. The weights are given by equation~(\ref{eq:wt}) and the error covariance reduces to
\begin{equation}
    \cov_{ij}^{\rm G}=\frac{1}{\sum_{g_{i}}\tilde{w}_{g_{i}}} \delta_{ij}~.
    \label{eq:covG_obs}
\end{equation}
In this work, we present results for two different observation times, namely medium and long which correspond to $t_{\rm obs}=1024$ and $10000$ hours respectively. The system noise contribution is Gaussian and it decreases with increasing  observation time $t_{\rm obs}$. We expect the error covariance  (equation \ref{eq:cov_obs}) to approach the Gaussian prediction (equation \ref{eq:covG_obs})  for small and also moderate $t_{\rm obs}$ where it is system noise dominated, whereas the non-Gaussianity is relatively more important for longer observation times. 

\subsubsection{Foregrounds}\label{sec:fgd}
The low-frequency radio sky is dominated by  Galactic and the extra-galactic foregrounds  which are several orders of magnitude brighter than the expected EoR 21-cm signal (e.g. \citealt{Ali_2008,Abhik_2012,Paciga_2013,Beardsley_2016,Barry_2019,Li_2019}). The foregrounds contaminated $\kk$ modes are largely expected  to be restricted within a wedge shape region in the $(\kk_\perp, k_\parallel)$ plane \citep{Datta_2010}, the boundary of this wedge being given by \citep{Morales_2012} 
\begin{equation}
k_{\parallel} = \left[\frac{r_c ~\sin(\theta_{\rm L})}{r^\prime_c ~\nu_c} \right] \times k_{\perp}
\label{eq:wedge}
\end{equation}
where $\theta_{\rm L}$ is the maximum angle on the sky (relative to the pointing direction of the telescope) from which foregrounds contaminate the signal. The $(\kk_{\perp},k_\parallel)$ modes outside this foreground wedge are expected to be free of foreground contamination, and only these $\kk$ modes can be used for estimating the 21-cm PS.  In \citetalias{shaw_2019}, we have studied the impact of foregrounds on the 21-cm PS error covariance estimates considering three different foreground contamination scenarios. The first is the `Optimistic' scenario where the foregrounds are assumed to be perfectly modelled and completely removed whereby the entire $(\kk_{\perp},k_\parallel)$ plane can be used for estimating the 21-cm PS. Next are the `Moderate' and the `Pessimistic' scenarios where we assume that there is a substantial foreground contamination coming from the sky within an angle $\theta_{\rm L}=3\times {\rm FWHM}/2$ and $90^\circ$ respectively. We discard the foreground contaminated modes from the estimation of the bin-averaged 21-cm PS and its error covariance. The volume of the discarded $\kk$ modes varies depending on the observing redshift as well as on the foreground scenario.

\citetalias{shaw_2019} presents detailed predictions for the error covariance matrix for the three different foreground scenarios mentioned above. As we move from the Optimistic to the Moderate and then the Pesimistic scenario, the region of  $(\kk_{\perp},k_\parallel)$ plane available for estimating the 21-cm PS gets smaller, and the SNR also falls. It is important to note that the error estimate approaches the Gaussian predictions as the SNR goes down, however the non-Gaussian contributions are important at high SNR \citep{Mondal_2015}. 

\subsection{Power spectrum derivatives}
\label{sub:derv}

\begin{figure*}
\centering
\includegraphics[width=0.95\textwidth]{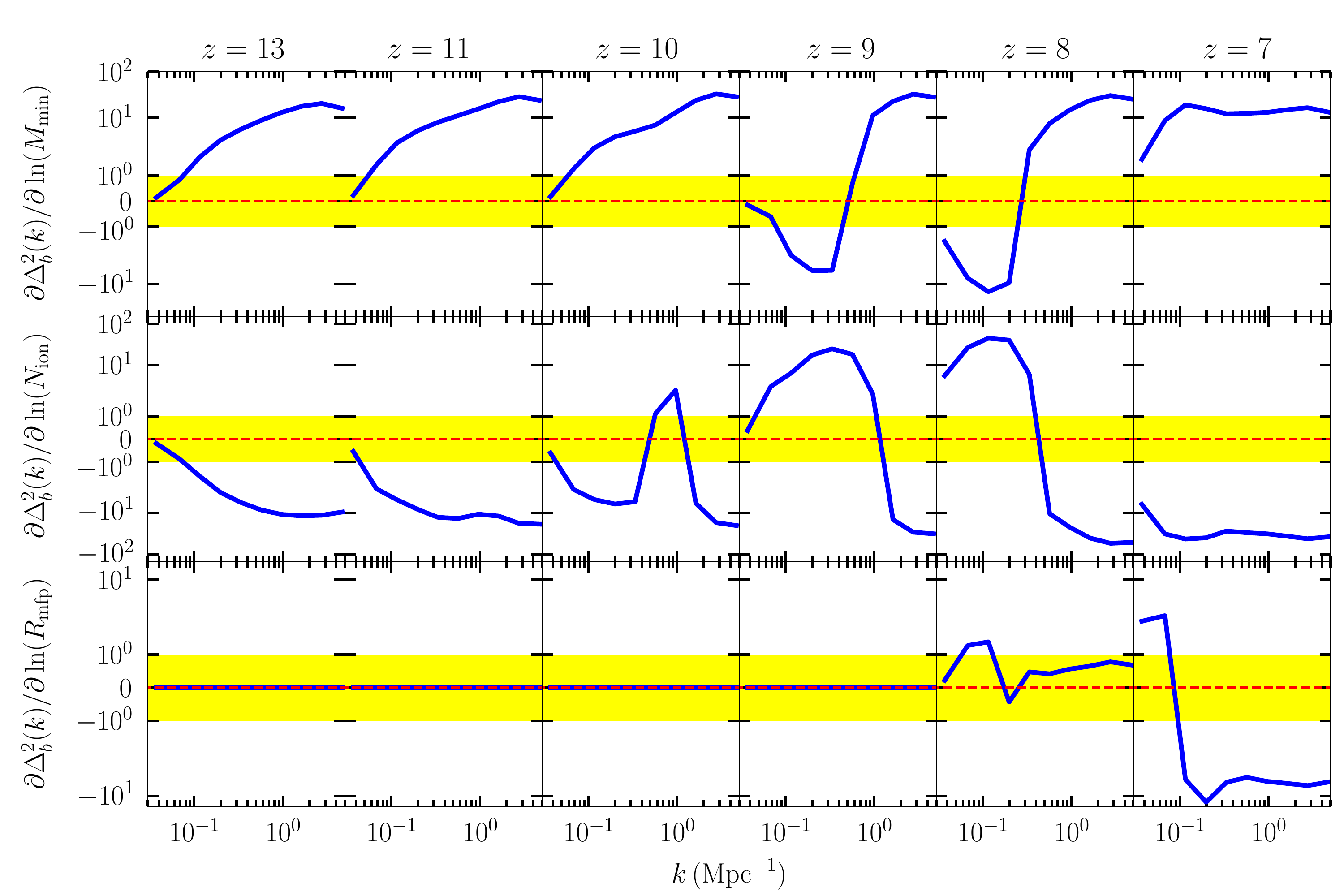}
\caption{The derivatives of $\Delta_{\rm b}^2(k)$ with respect to the three log-parameters computed at six different redshifts. The red dotted line is the zero reference line and the yellow shade demarcates the region where the y-scale is linear and logarithmic otherwise.}
\label{fig:derv}
\end{figure*}

In order to calculate the Fisher matrix $F_{\alpha \beta}$ (equation \ref{eq:fisher}), we need the 21-cm PS error covariance $\cov_{ij}$ and $\partial \bar{P}(k_i)/\partial{\bm q}$ which is the partial derivatives of the bin-averaged 21-cm PS $\bar{P}(k_i)$ with respect to the parameters ${\bm q}$. These partial derivatives behave like weights that modulate the contribution of the 21-cm PS error covariance $\cov_{ij}$ to the Fisher matrix $F_{\alpha \beta}$. Here it is convenient to use $q_{\alpha}=[\ln(\Mmin),\,\ln(\Nion),\,\ln(\Rmfp)]$ as the parameters so that we can directly interpret $\Delta q_1=\Delta \Mmin/\Mmin,\cdots$ as the fractional errors in the respective astrophysical parameters. 

We compute the partial derivatives of the 21-cm PS at the fiducial parameter values $q_{\alpha_{\rm o}}=[1.09\times 10^9~{\rm M}_\odot,~23.21,~20\mpc]$ using a numerical six-point derivative formula
\begin{equation}
\frac{\partial \bar{P}(k_i)}{\partial q_\alpha} = \frac{{\mathcal X}(4h_\alpha)-40{\mathcal X}(2h_\alpha)+256 {\mathcal X}(h_\alpha)}{360~h_\alpha}~,
\label{eq:derv}
\end{equation}
where ${\mathcal X}(N h_\alpha)\equiv [\bar{P}(k_i \mid q_{\alpha_{\rm o}}+ N h_\alpha)-\bar{P}(k_i \mid q_{\alpha_{\rm o}}-N h_\alpha)]$ and $N=(1,~2,~4)$. In the above equation, $h_\alpha=[0.042,~0.022,~0.125]$ denotes the step size corresponding to the three parameters $q_\alpha=[\ln(\Mmin),\,\ln(\Nion),\,\ln(\Rmfp)]$ respectively. To evaluate equation (\ref{eq:derv}) we have run  the reionization simulations with the parameter values $q_{\alpha}=q_{\alpha_{\rm o}} \pm N h_\alpha$. We individually vary each parameter keeping the values of the other parameters fixed at their fiducial values to estimate the partial derivatives. We have used the 21-cm PS estimated from these simulations to calculate ${\mathcal X}(N h_\alpha)$ and evaluate  the partial derivatives of the 21-cm PS.

The fiducial value of $\Mmin$ ($=1.09\times 10^9~{\rm M}_\odot$) in our reionization simulations is the same as the mass of the smallest halos from our \textit{N}-body simulations (section \ref{sec:sim}). We however require halos with  masses smaller than $1.09\times 10^9~{\rm M}_\odot$  for estimating the 21-cm PS derivatives with respect to $\Mmin$. For this purpose, we run a higher resolution \textit{N}-body simulation which has a grid spacing of $0.0525\mpc$ maintaining the box size same as the earlier simulations. These simulation has a higher mass-resolution and the smallest resolved halo has a mass of $4.59\times 10^8 ~{\rm M}_\odot$. However, the 21-cm brightness temperature fluctuations were generated on the same grid (same spatial resolution) as in our fiducial reionization simulations to maintain the $k$ binning of all simulations identical.

The different panels in Figure \ref{fig:derv} show the derivatives of the dimensionless bin-averaged 21-cm PS $\Delta_{\rm b}^2(k)$ as a function of wave number $k$. The panels are arranged in a way where the three different rows correspond to the three different parameters and the different columns correspond to the six different redshift  considered in our analysis. The yellow shade demarcates the region where the scale of the vertical axis is linear. The scale outside the yellow shaded region is logarithmic.

The top row of Figure \ref{fig:derv} shows $\partial \Delta_{\rm b}^2(k_i)/\partial \ln(\Mmin)$ as a function of $k$. We see that this  is positive for all $k$ during the initial stages of reionization ($z \ge 10$) and also at the very end stage of  reionization ($z=7$). In the intermediate stage $(8 \leq z <10)$ we find that $\partial \Delta_{\rm b}^2(k_i)/\partial \ln(\Mmin)$ is positive at large $k$, however this is negative at small $k$. We can interpret the behaviour of this derivative at $z \ge 10$ in terms of Figure \ref{fig:pk_tri} which shows the evolution of the 21-cm PS with $z$. Note that increasing $\Mmin$ reduces the number of ionization sources and delays reionization, the effect is similar to considering a higher $z$. In Figure \ref{fig:pk_tri} we see that at all $k$ the 21-cm PS drops with decreasing $z$ for $z \ge 10$,  this  explains the positive value of the derivative in this $z$ range. At $z \le 10$ the typical ionized bubble size is imprinted in the 21-cm PS 
(eq. 22 of \citealt{Bharadwaj_2005}) and also its derivatives. The $k$ value where $\partial \Delta_{\rm b}^2(k_i)/\partial \ln(\Mmin)$ changes sign approximately corresponds to the bubble radius at the particular redshift. The typical bubble size is comparable to the simulation box at $z=7$ where the derivative is positive everywhere.

The middle row of Figure \ref{fig:derv} shows  $\partial \Delta_{\rm b}^2(k_i)/\partial \ln(\Nion)$ as a function of $k$. We see that this is negative for all values of $k$ in the early stages of reionization $(z >10)$ and also at the very end stage of reionization ($z=7$). At $z=10$ this derivative is negative for all $k$, except for a small positive kink  around $k =1\impc$. In the intermediate  stage $(8 \leq z <10)$ we find that $\partial \Delta_{\rm b}^2(k_i)/\partial \ln(\Mmin)$ is negative at large $k$, however this is positive at small $k$. Overall we see that $\partial \Delta_{\rm b}^2(k_i)/\partial \ln(\Nion)$ is very similar to $\partial \Delta_{\rm b}^2(k_i)/\partial \ln(\Mmin)$, except that the sign is reversed. We can interpret this by noting that increasing $\Nion$ is akin to lowering $\Mmin$ in that  both of these hasten reionization. The extra kink around $k =1\impc$ seen here at $z=10$ is related to the typical size of the ionized bubble at this redshift.

The bottom row of Figure \ref{fig:derv} shows $\partial \Delta_{\rm b}^2(k_i)/\partial \ln(\Rmfp)$. We find that the EoR 21-cm PS in our simulations is not sensitive at all ($\partial \Delta_{\rm b}^2(k_i)/\partial \ln(\Rmfp)=0$) to the mean free path of the ionizing photons at $z\geq 10$, and the derivatives are very small ($\sim 10^{-7}-10^{-4}$) at $z=9$. We believe that this is due to the fact that at $z\geq 9$ the typical sizes of the \HII bubbles is smaller than the fiducial value of $\Rmfp ~(=20\mpc)$ in our simulations. Considering $\partial \Delta_{\rm b}^2(k_i)/\partial \ln(\Nion)$, the $k$ values corresponding to the kink seen at $z=10$ and the sign change seen at $z=9$ provide estimates of the typical bubble size at the respective redshifts. We see that these estimates both confirm that the typical bubble size is smaller than $20 \mpc$. The 21-cm PS does depend on $\Rmfp$ during the later stages of reionization ($z\leq 8$). We see that $\partial \Delta_{\rm b}^2(k_i)/\partial \ln(\Rmfp)$ is particularly large at $z=7$ where it is positive at small $k$ and negative at large $k$. The $k$ value corresponding to the transition approximately matches the fiducial value of $\Rmfp = 20\mpc$.

We note that the 21-cm PS derivatives obtained in our analysis are qualitatively very similar to the results in \citet{Pober_2014} who have considered such derivatives in an earlier work. However, it is necessary to note that their reionization model and the fiducial parameter values are quite different from the ones used here. Our results, though qualitatively similar to \citet{Pober_2014}, differ in the quantitative details.

We use the numerically obtained partial derivatives $\partial \bar{P}(k_i)/\partial{\bm q}$ and the inverse of PS error covariance matrix $\cov_{ij}$ to evaluate the Fisher matrix $F_{\alpha \beta}$ (equation~\ref{eq:fisher}) of our model parameters. The inverse of the Fisher matrix provides the corresponding parameter error covariance $\covp_{\alpha\beta}$ (equation \ref{eq:cr}), the fractional errors in the parameters $\Delta q_\alpha$ in our analysis. 

%======================================================================================%
\section{Results}\label{sec:res}

The question here is `How accurately can we estimate the parameters of our reionization model given a 21-cm power spectrum (PS) measurement?'. We quantify this using  $\covp_{\alpha \beta}$  which is the error covariance matrix for the model parameters, with $\covp_{\alpha \beta}$  here being calculated using equation~(\ref{eq:cr}) which relates it to the Fisher matrix  (equation~\ref{eq:fisher}). Non-Gaussian effects enter into our calculation of $\covp_{\alpha \beta}$ through the trispectrum that contributes to the 21-cm PS error covariance matrix $\cov_{ij}$ (equations~\ref{eq:cov_CV} and \ref{eq:cov_obs}). Our analysis particularly focuses on studying the impact of non-Gaussianity on error predictions for the reionization parameters. We find (Figure \ref{fig:derv}) that varying $\Rmfp$ has no effect on the 21-cm PS at $z\geq 10$ and therefore the analysis in this redshift range is restricted to only two parameters namely $\Mmin$ and $\Nion$ whereas we have considered three parameters ($\Rmfp,~ \Mmin,~\Nion$) at $z<10$. 

We have presented the results of our analysis in two stages. In the first stage, we only consider the signal without incorporating any of the observational effects and analyze how cosmic variance arising from the statistical uncertainties inherent to the signal affects parameter estimation. This allows us to study the effect of non-Gaussianity without reference to any particular instrument or observations. The results here are based on a simulation volume $V=[215.04\mpc]^3$. \citet{Mondal_I} have shown that this volume is large enough for the 21-cm PS to converge. Small differences are noted in values of trispectra when compared to a smaller simulation volume ($[150\mpc]^3$). However we do not expect trispectra to change much for the larger volumes, and assume the trispectrum would converge for the $k$ range of our interest. In the second stage, we introduce instrumental effects, and we make predictions specific to future observations with SKA-Low. We also incorporate the effects of foregrounds here. 

\subsection{Constraints considering Signal only}
\label{sub:res_sim}

\begin{figure}
\centering
\includegraphics[scale=0.5]{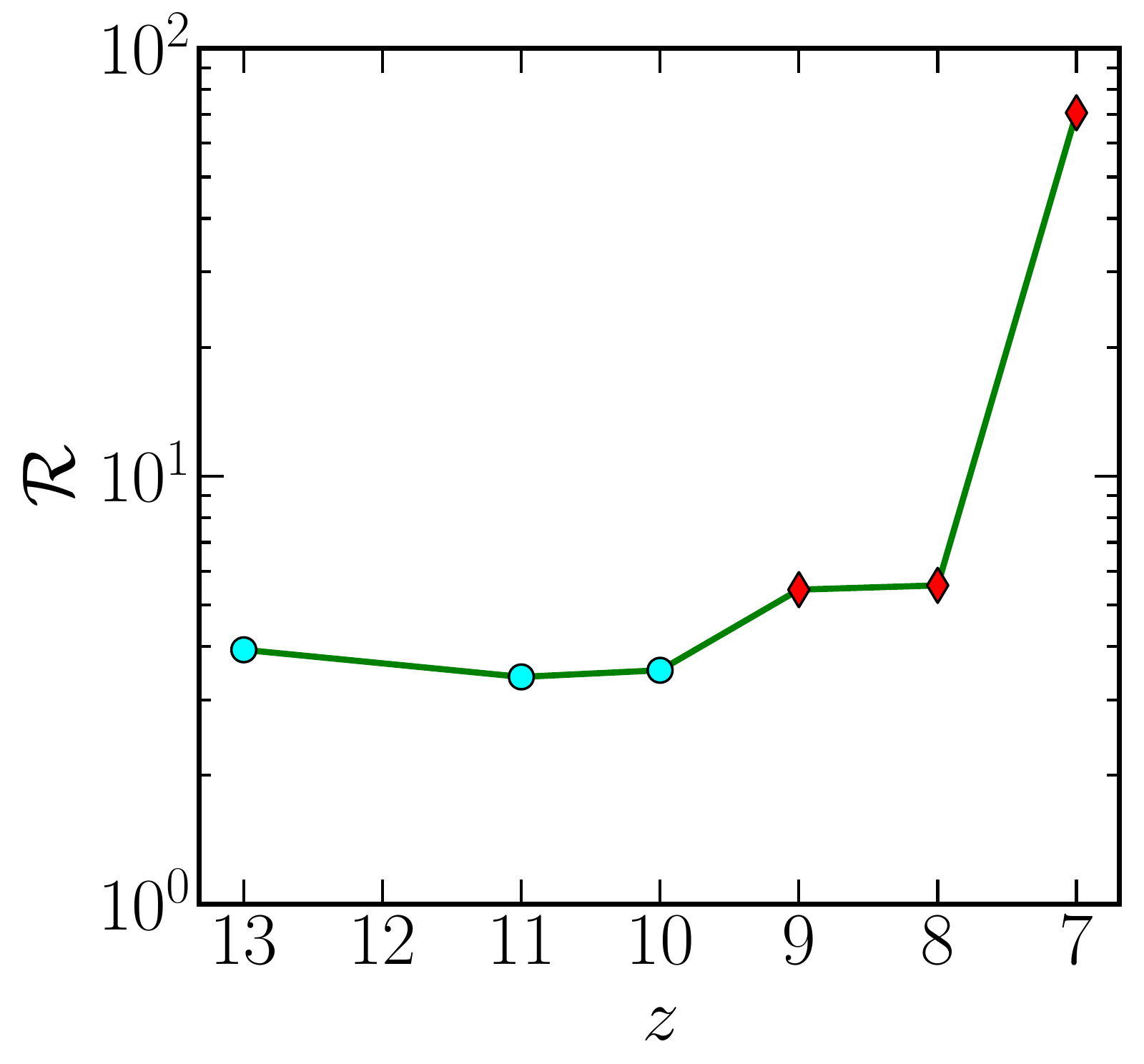}
\caption{The redshift evolution of the ratio $\mathcal{R}$. The 2D and 3D data points are denoted by circles and diamonds respectively. However, the constraints on $\Rmfp$ is weak at $z=9$, for which we consider 2D Fisher matrix in rest of our analysis.}
\label{fig:vol}
\end{figure}

\begin{figure*}
\includegraphics[scale=0.8]{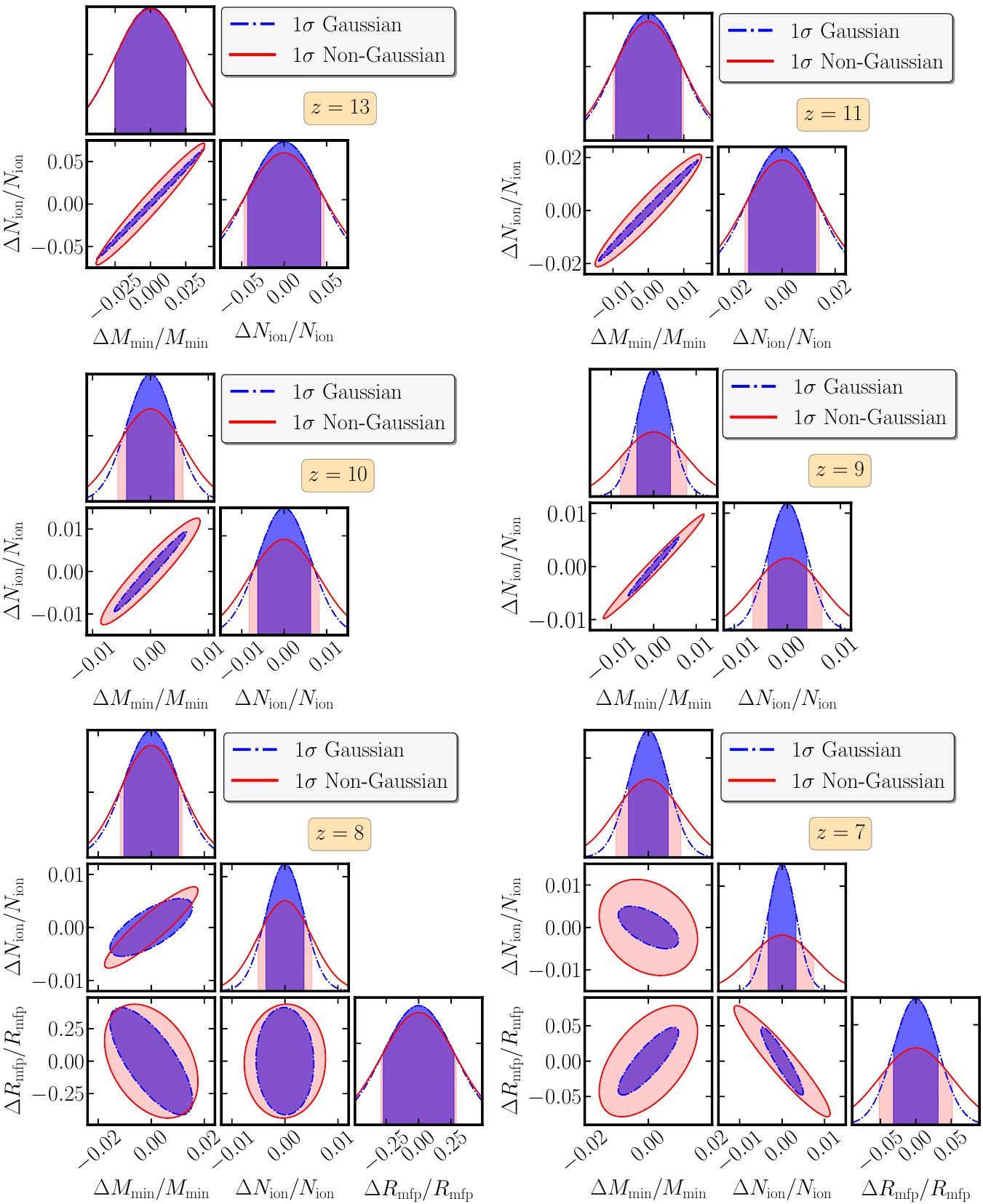}
\caption{The marginalized $1\sigma$ error ellipses and probability distribution of fractional errors in parameters considering only cosmic variance as a source of error in 21-cm PS.}
\label{fig:elp_cv}
\end{figure*}

We first consider the expected 21-cm signal alone without reference to any particular instrument. The errors here are due to 
the cosmic variance (CV) which arises from the limited volume and the statistical uncertainties inherent to the signal. 
Here we focus on the $1\sigma$ error ellipsoid in the three dimensional (3D) $\Mmin,\Nion,\Rmfp$ parameter space. As noted earlier, this reduces to a two dimensional (2D) ellipse in the $\Mmin,\Nion$ parameter space for $z \ge 10$. We find that the size and orientation of the error ellipsoid (ellipse) both change as reionization proceeds. Further, the size and orientation also change depending on whether we consider the Gaussian or non-Gaussian predictions. The volume (area) of the ellipsoid (ellipse) provides a broad quantitative measure of the errors in parameter estimation.  In order to quantify how non-Gaussianity affects parameter estimation we consider $\mathcal{R}$ which is the ratio of the non-Gaussian prediction to the Gaussian prediction for the volume (area). In Figure \ref{fig:vol} we see that $\mathcal{R}$ has values in the range $3-4$ for $z\geq 10$ where $\covp_{\alpha \beta}$ is 2D. These redshifts correspond to the initial stages of reionization where $\xb \geq 0.86$. We see that non-Gaussianity has a noticeable effect on parameter estimation even during these early stages of reionization where the area of the error ellipse is predicted to be several times larger than the Gaussian predictions. The non-Gaussianity increases as reionization proceeds, and the ratio $\mathcal{R}$ rises to values around $\sim 6$ for $8 \leq z \leq 9$ where $0.50 \leq \xb \leq 0.73$ and $\covp_{\alpha \beta}$ is 3D. We see that the effect of non-Gaussianity increases sharply at $z=7$ ($\xb\simeq0.15$) where $ \mathcal{R} \simeq 70$. The non-Gaussian effects are very important for the error predictions during the final stages of reionization.

Figure \ref{fig:elp_cv} provides a detailed analysis of the error estimates across the redshift range of our interest. We first consider $z=13$ for which we have three panels of which the lower left panel shows the $1\sigma$ error ellipses in the $\Mmin,\Nion$ plane. We find that the major axis of both the Gaussian and the non-Gaussian error ellipses have approximately equal positive slope of $\approx 60^{\circ}$  which indicates a positive correlation among $\DM$ and $\DN$. We  notice that the major axis of the non-Gaussian error ellipse is only slightly larger than that of the Gaussian, however the the minor axes is $3.67$ times larger as also reflected in the value of $\mathcal{R}$ (Figure \ref{fig:vol}). Considering the panels which show the respective marginalized one dimensional (1D) errors, we see that $\DM$ and $\DN$ are $0.0253$ and $0.0469$ respectively, with very little difference between the Gaussian and non-Gaussian predictions. The two marginalized errors are related to the projections of the 2D error ellipse on the respective axes. Here the $\approx 60^{\circ}$ slope of the ellipse causes both the  $\DM$ and $\DN$ projections to be determined by the major axis whose value does not differ much for the Gaussian and non-Gaussian predictions. 

The results at $z=11$ are very similar to those at $z=13$ except that the errors are now smaller with $\DM$ and $\DN$ having  values $0.0099$ and $0.0140$ respectively. The slope of both the major axes are around $55^\circ$ which is less with respect to that for $z=13$.  Here the ratio of non-Gaussian to the Gaussian minor axis is $3.13$ which is consistent with the value of $\mathcal{R}$. The differences between the Gaussian and non-Gaussian 1D errors are a little more pronounced in comparison to $z=13$; however, the differences are still rather small. 

\begin{table*}
\centering
\begin{tabular}{|c|ccc|ccc|ccc|}
\hline
\multirow{2}{*}{$z$} & \multicolumn{3}{c|}{$(\DM) ~\times10^{-2}$}      & \multicolumn{3}{c|}{$(\DN) ~ \times10^{-2}$}      & \multicolumn{3}{c|}{$(\DR)~ \times 10^{-2}$}      \\ \cline{2-10} 
                   & \multicolumn{1}{c}{Non-Gaussian} & Gaussian & $\Delta(\%)$ & \multicolumn{1}{c}{Non-Gaussian} & Gaussian & $\Delta(\%)$ & \multicolumn{1}{c}{Non-Gaussian} & Gaussian & $\Delta(\%)$ \\ \hline
\rowcolor[HTML]{EFEFEF} $13$ & $2.53$ & $2.51$ & $1$ & $4.69$ & $4.36$ & $10$ & $-$ & $-$ & $-$ \\
                        $11$ & $0.99$ & $0.93$ & $7$ & $1.40$ & $1.26$ & $11$ & $-$ & $-$ & $-$ \\
\rowcolor[HTML]{EFEFEF} $10$ & $0.57$ & $0.41$ & $37$ & $0.83$ & $0.62$ & $33$ & $-$ & $-$ & $-$ \\
                        $9$ & $0.79$ & $0.41$ & $93$ & $0.65$ & $0.37$ & $75$ & $-$ & $-$ & $-$ \\
\rowcolor[HTML]{EFEFEF} $8$ & $1.16$ & $1.03$ & $13$ & $0.51$ & $0.36$ & $41$ & $28.90$ & $27.25$ & $6$ \\
                        $7$ & $1.02$ & $0.62$ & $63$ & $0.75$ & $0.33$ & $27$ & $5.20$ & $3.16$ & $64$ \\ \hline
\end{tabular}
\caption{The $1\sigma$ fractional errors (first two sub-columns) for each inferred parameter considering only the cosmic variance as a source of error in the measured 21-cm PS. Here $\Delta(\%)$ (third sub-column) is the percentage deviation of the non-Gaussian predictions from the Gaussian ones.}
\label{tab:1}
\end{table*}

The impact of non-Gaussianity increases at $z=10$. Here also the Gaussian and non-Gaussian 2D error ellipse are aligned, and both  have a slope of $\approx  56^{\circ}$. For the non-Gaussian ellipse the major and minor axes are respectively $1.33$ and $2.64$ times the Gaussian values with $\mathcal{R} \approx 3.52$. The 1D non-Gaussian error predictions for $\DM$ and $\DN$ are $0.0057$ and $0.0083$ respectively. At this redshift, we see that non-Gaussianity has a considerable effect on the marginalized 1D error predictions with $\DM$ and $\DN$ being respectively around $37 \%$ and $33 \%$ larger than the Gaussian predictions.

The error covariance matrix $\covp_{\alpha \beta}$ is 3D for $z\leq 9$. However at $z=9$, the errors for $\Rmfp$ are extremely large compared to the errors in the other parameters and we have marginalized over $\Rmfp$ leading to a 2D analysis at this redshift. The Gaussian and non-Gaussian error ellipses are aligned and have a slope of approximately $48^\circ$. We note that the area of the Gaussian ellipse is smallest for this redshift, and the non-Gaussian major and minor axes are respectively $1.85$ and $2.00$ times those of the Gaussian. Considering the 1D marginalized errors, the non-Gaussian estimates predict that $\DM$ and $\DN$ are $\sim 0.0079$ and $0.0065$ respectively. Here $\DM$ is slightly larger than for $z=10$. We also note that non-Gaussianity has a considerable effect on the 1D marginalized error predictions at this $z$, and these predictions are $93 \%$ and $75 \%$ in excess of the Gaussian predictions for $\DM$ and $\DN$ respectively. 
 
We next consider $z=8$ for which we present a full 3D analysis. We first consider the left panel of the middle row which shows the 2D error ellipse in the $\Mmin,\Nion$ plane where  we have marginalized over the third parameter $\Rmfp$. Considering the non-Gaussian and Gaussian ellipses, we see that the major axes are not exactly aligned, these being respectively tilted at  $22^\circ$ and $15^\circ$ with respect to the horizontal. The non-Gaussian major and minor axes are respectively $1.18$ and $0.79$ times the Gaussian values, interestingly here the area of the non-Gaussian ellipse is smaller than that of the Gaussian. We next consider the bottom row where the left and middle panels respectively show the $\Mmin,\Rmfp$ and $\Nion,\Rmfp$ error ellipses with the third parameter is marginalized. We see that the errors in $\Rmfp$ are considerably bigger compared to those in the other two parameters, and the ellipses are both nearly upright with slopes in the range $89^\circ-92^\circ$. Our results indicate that the errors in $\Rmfp$ are largely uncorrelated with those in the other two parameters which are positively correlated amongst themselves. Comparing the non-Gaussian to the Gaussian error ellipses, the major axes are comparable but the minor axes are $1.54$ and $1.41$ times larger in the left and right panels respectively. We next consider the 1D marginalized errors where $\DM$, $\DN$ and $\DR$ have values $0.0116,~0.0051~{\rm and}~0.2890$ respectively. The non-Gaussian predictions are $13 \%$ and $41 \%$ larger than the corresponding Gaussian predictions for $\DM$ and $\DN$ respectively. However we hardly observe very noticeable difference between the non-Gaussian and Gaussian predictions for $\DR$.

Considering $z=7$ we see that the results are quite different from those at earlier redshifts, the effect of non-Gaussianity is also most pronounced at this redshift. Considering the middle row left panel, we find that the major axis of the $\Mmin,\Nion$ non-Gaussian and Gaussian error ellipses are both at $\approx 160^\circ$ to the horizontal which indicates an anticorrelation between the errors in these two parameters. The non-Gaussian ellipse is quite a bit larger and the major and minor axes are respectively $1.59$ and $2.80$ times those of the Gaussian ellipse. Considering the bottom row we see that the errors in $\Rmfp$ are relatively large compared to those in the other two parameters, and the $\Mmin,\Rmfp$ (left) and the $\Nion,\Rmfp$ (middle) error ellipses both have their major axes nearly upright. For the former, the slopes of the non-Gaussian and Gaussian major axes are $84^\circ$ and $80^\circ$ respectively which indicates a mild correlation in the errors. The non-Gaussian major and minor axes are respectively $1.63$ and $2.60$ times larger than the Gaussian predictions. For the latter ($\Nion,\Rmfp$)  we see that the non-Gaussian and Gaussian major axes respectively have slopes of $98^\circ$ and $95^\circ$ with respect to the horizontal which indicates mild anticorrelations between the errors. The non-Gaussian major and minor axes are respectively  $1.65$ and $2.08$ times the Gaussian predictions. Considering the 1D predictions we find that the non-Gaussian predictions for $\DM$ and $\DR$ are respectively $\sim 0.0102$ and $0.0520$ which are smaller than those at $z=8$, however $\DN$ which is $0.0075$ is slightly larger. The non-Gaussian predictions for $\DM$, $\DN$ and $\DR$ are $\sim 63 \%$, $127 \%$ and $65 \%$ larger than the respective Gaussian predictions.

Considering all the panels in Figure \ref{fig:elp_cv} together we note that the orientation of the error ellipses which quantify the nature of correlations between the errors of various pairs of parameters is nearly the same whether we consider the non-Gaussian or Gaussian predictions. Further, the orientation also does not change much at $z > 8$. We, however, notice changes in the ellipse orientations at $z=7$ and $8$. The non-Gaussianity causes the area of the error ellipses to increase, this is also reflected in the marginalized 1D errors. Table~\ref{tab:1} summarizes the 1D marginalized errors (both non-Gaussian and Gaussian) across the entire redshift range considered here. The minima of the non-Gaussian predictions of $\DM$, $\DN$ and $\DR$ occurs at $z=10,~8,~{\rm and}~7$ respectively. However the minima of the corresponding Gaussian error predictions are respectively at $z=9,~7,~{\rm and}~7$. An earlier study \citep{Mondal_II} shows that at small length-scales ($k=2.75 \impc$) the 21-cm signal becomes increasingly non-Gaussian as reionization proceeds. The same is also true at intermediate ($k = 0.57 \impc$) and large ($k = 0.12 \impc$) length-scales except that there is a dip at $z=8$ ($\xb=0.5$) beyond which it increases again. We see that the differences between the non-Gaussian and Gaussian parameter error predictions shows a behaviour similar to that seen at intermediate and large scales where the differences increase as reionization proceeds except for a dip at $z=8$ beyond which it increases again. 

Note that the results presented in this subsection are particular to the aforementioned simulation volume $V=[215.04\mpc]^3$. \citet{Mondal_I} have explicitly shown that the cosmic variance $\cov_{ij} \propto V^{-1}$, provided the  bin boundaries are fixed in $\kk$ space. This indicates an overall decrease in the cosmic variance if the volume is increased. However, the ratio of the Gaussian predictions to the non-Gaussian predictions remains invariant as both have exactly the same volume dependence. This also carries over to the error estimates for the individual reionization parameters. In addition to this, we also have some additional small $k$ modes if the volume is increased. This is an additional contribution which further reduces the error predictions. Whether these additional modes increase or decrease the relative effect of non-Gaussianity depends on the trispectrum at these $k$ values.

\subsection{Constraints considering SKA-Low observations}
In real observations, the error variance of the observed EoR 21-cm PS will have contributions from various other sources such as system noise and calibration errors etc. In this analysis, we only consider the cosmic variance and the Gaussian system noise contributions to the error in the measured 21-cm PS. The system noise only affects the diagonal elements of the 21-cm PS error covariance $\cov_{ij}$, the off-diagonal terms remain unaffected. As the noise PS varies inversely with the observation time $t_{\rm obs}$, the system noise contribution to $\cov_{ii}$ varies as $t^{-2}_{\rm obs}$ whereas the CV contribution is  independent of $t_{\rm obs}$. As a consequence of this, the impact of non-Gaussianity becomes more pronounced in the $\cov_{ij}$ if we observe for a longer time \citep{shaw_2019}. In the present analysis, we consider two different cases namely a medium observation time ($t_{\rm obs}=1024$ hours) and a very long observation time ($t_{\rm obs}=10000$ hours). We also present results considering infinitely long observation time, \textit{i.e.} $t_{\rm obs}\rightarrow \infty$ where the $\cov_{ij}$ will hit the CV limit. We also consider the foreground effects, and present our results for the three foreground scenarios namely Optimistic, Moderate and Pessimistic that have been discussed in Section \ref{sec:fgd}. For the present analysis, we have combined the Fisher matrix from all the six redshifts ($z=13,~11,~10,~9,~8,~7$) to improve the constraints on the three reionization parameters. Note that for each redshift the SKA-Low observational volume is larger than the simulation volume, we have accounted for this in the error covariance matrices (equation~\ref{eq:cov_obs}).

We first consider the full 3D error ellipsoids for which Table~\ref{tab:2} lists the values of $\mathcal{R}$ for the different foreground scenarios and the two observation times considered here, the CV values are also shown for reference. Note that the CV limit corresponds to the maximum value of $\mathcal{R}$ that can be achieved for any particular  foreground scenario.  We see that the values of $\mathcal{R}$ are $132.68$, $19.23$ and $6.95$ for the Optimistic, Moderate and Pessimistic foreground scenarios respectively. An earlier study shows that the non-Gaussian effects become progressively more important as larger number of $\kk$ modes are combined to increase the SNR \citep{Mondal_2015}, we see that this is also manifested here. The fact that increasingly larger number of $\kk$ modes have to be discarded for foreground avoidance as we go from the Optimistic to Moderate and Pessimistic scenarios is reflected in the behaviour of $\mathcal{R}$. The value of $\mathcal{R}$ falls drastically from the Optimistic to Moderate, the drop from Moderate to Pessimistic is not so severe.

We see that including the Gaussian system noise considerably reduces the effect of the non-Gaussianity in the 21-cm signal. The values of $\mathcal{R}$ fall to $1.45$ and $2.16$ for $t_{\rm obs}=1024$ and $10000$ hours respectively in the Optimistic scenario. Interestingly the non-Gaussian effects become relatively more important in the Moderate scenario where the values of $\mathcal{R}$ increase with respect to the Optimistic scenario for both $t_{\rm obs}=1024$ and $10000$ hours. This happens because the large $k$ bins that are system noise dominated are discarded due to the foreground contamination in the Moderate scenario \citep{shaw_2019}. The remaining intermediate and small $k$ bins, where there is a substantial trispectrum contribution, causes the effect of non-Gaussianity to increase relative to the Optimistic scenario. For the Pessimistic scenario, the values of $\mathcal{R}$ drop again but they are slightly larger than those for the Optimistic scenario.

\begin{table}
\centering
\begin{tabular}{c c c c}
\hline
$t_{\rm obs} \rightarrow$ & $1024$ hours & $10000$ hours & CV \\ \hline \hline
Optimistic & $1.45$ & $2.16$ & $132.68$ \\
Moderate & $2.42$ & $4.14$ & $19.23$ \\
Pessimistic & $1.65$ & $2.67$ & $6.95$ \\ \hline
\end{tabular}
\caption{The variation of the ratio $\mathcal{R}$ with observation time for the three foreground scenarios.}
\label{tab:2}
\end{table}

\begin{figure*}
\includegraphics[scale=0.27]{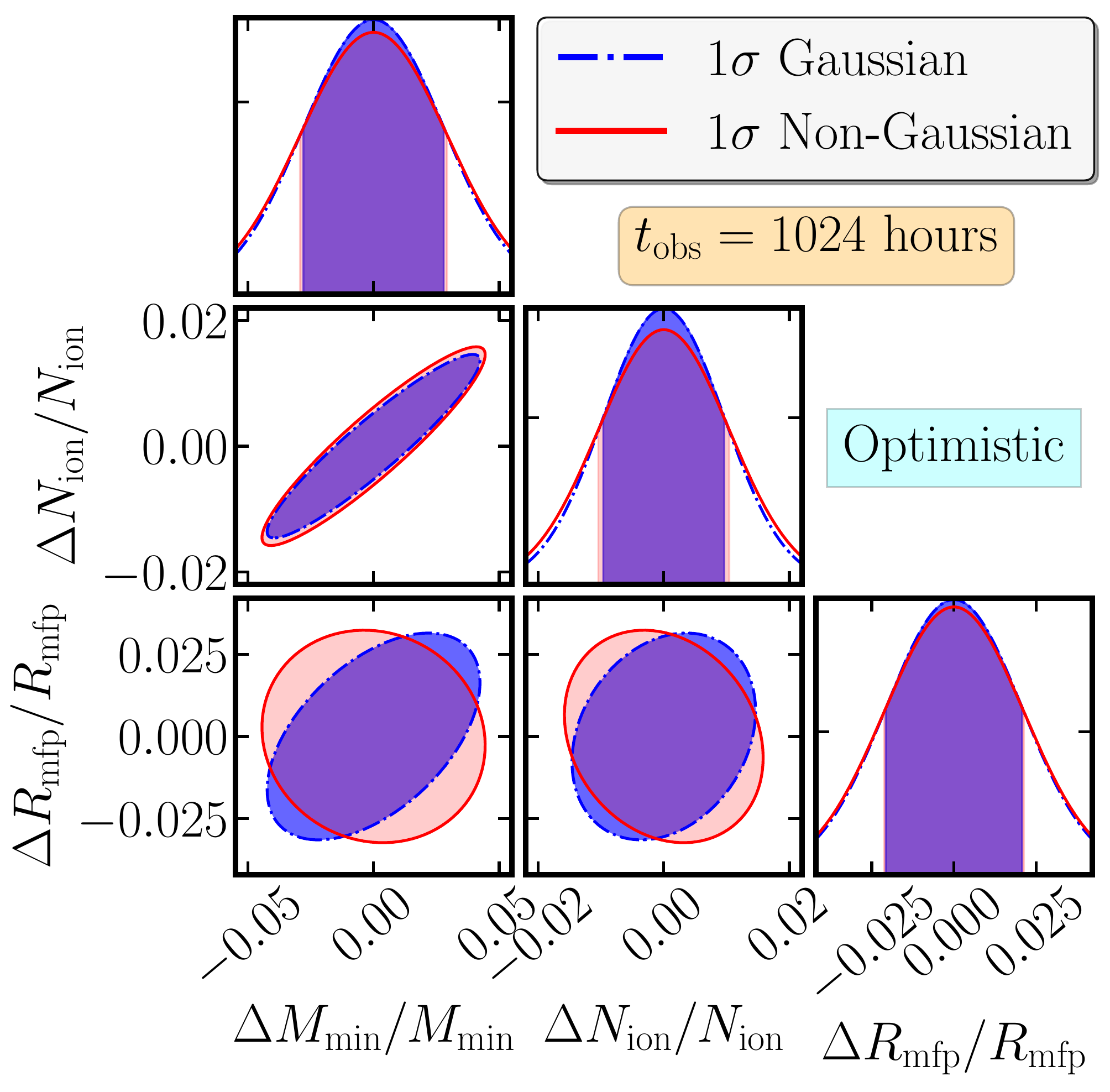}
\includegraphics[scale=0.27]{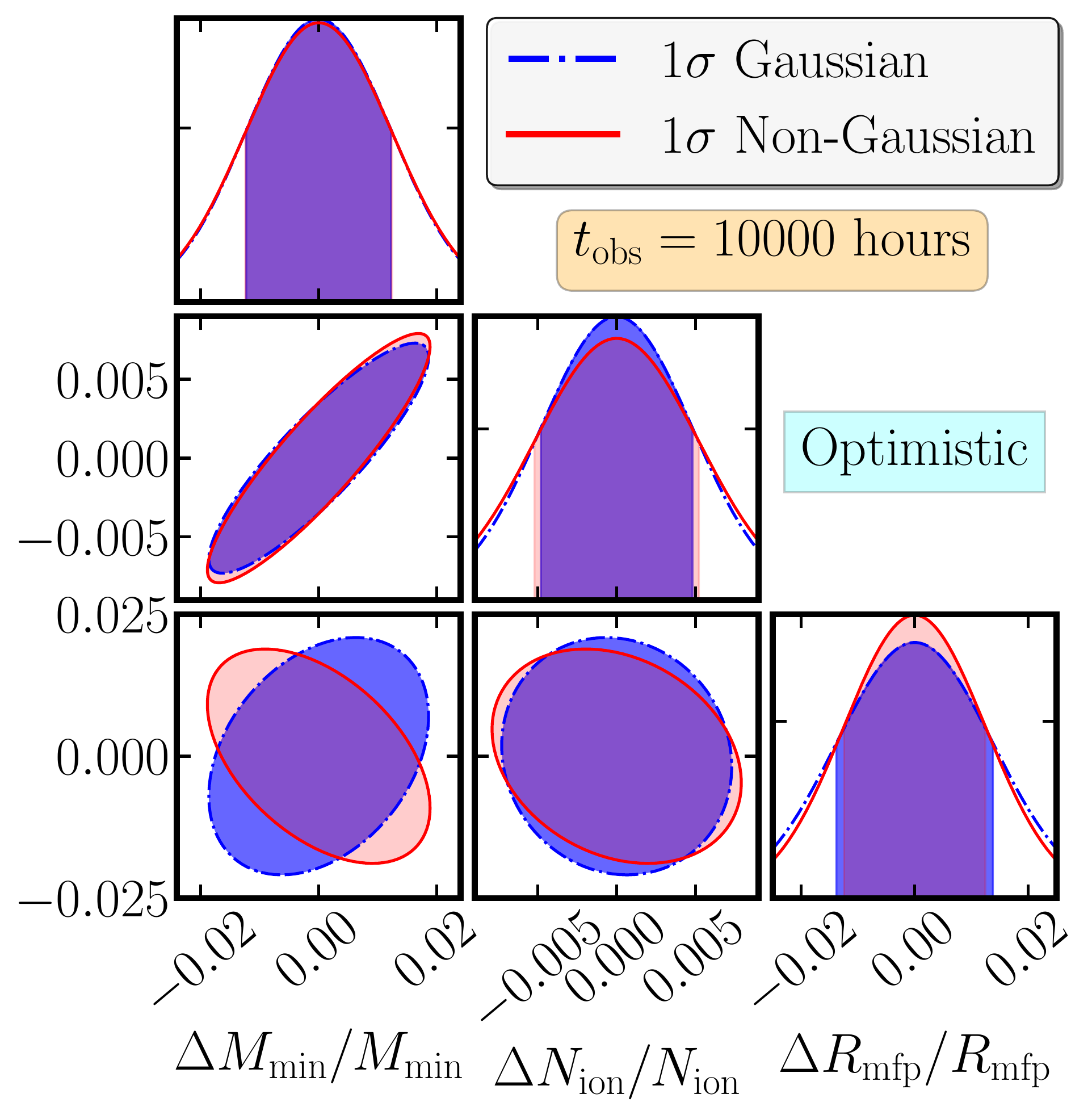}
\includegraphics[scale=0.27]{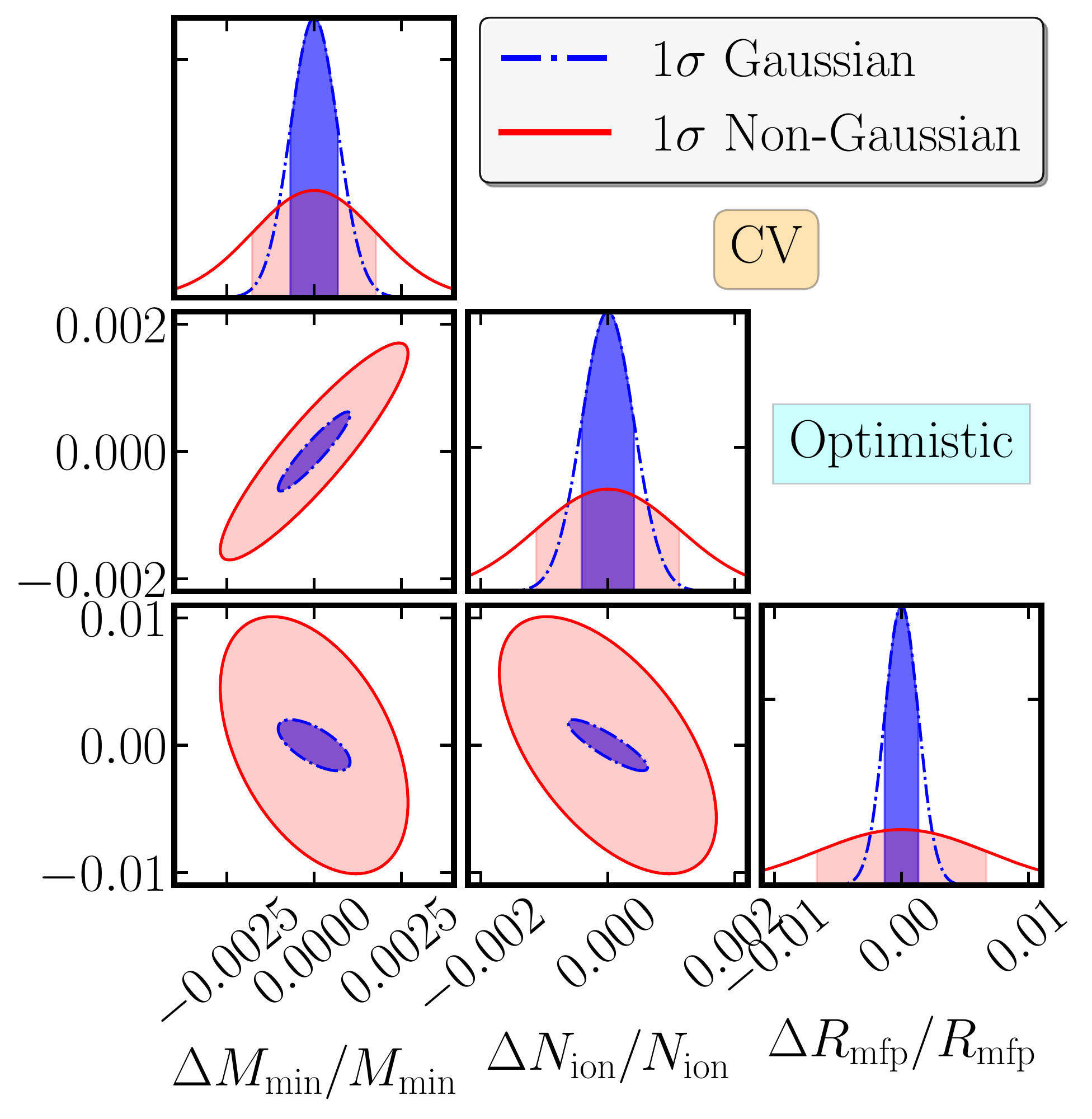}
\caption{The marginalized $1 \sigma$ error ellipses and 1D distribution of fractional errors in parameters for $t_{\rm obs}=1024$ hours (Left), $10000$ hours (Middle) and CV (Right) considering Optimistic foreground scenario. This predictions are obtained after combining Fisher matrices for all the six redshift slices.}
\label{fig:elp_obs_opt}
\end{figure*}

\begin{figure*}
\includegraphics[scale=0.27]{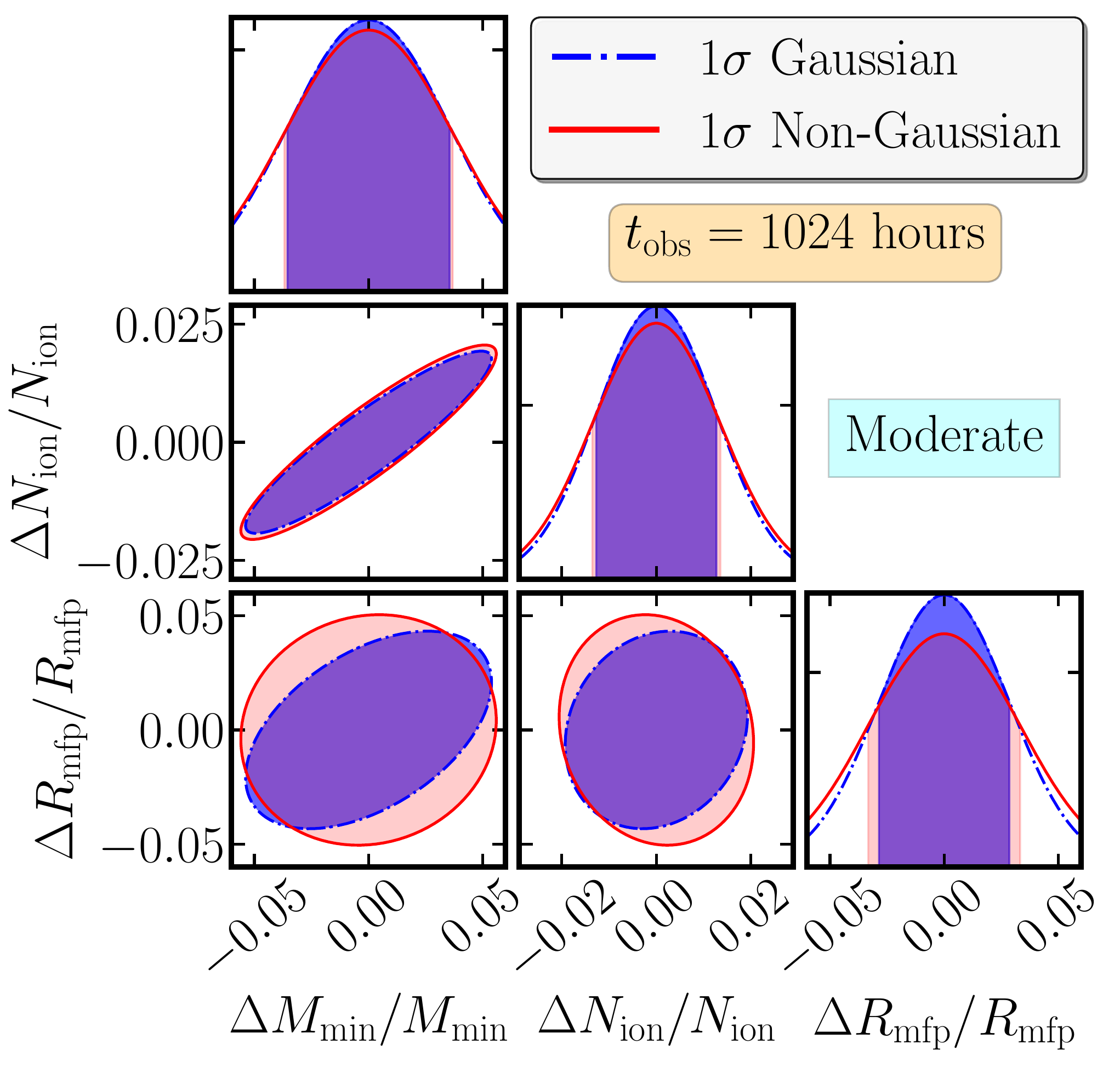}
\includegraphics[scale=0.27]{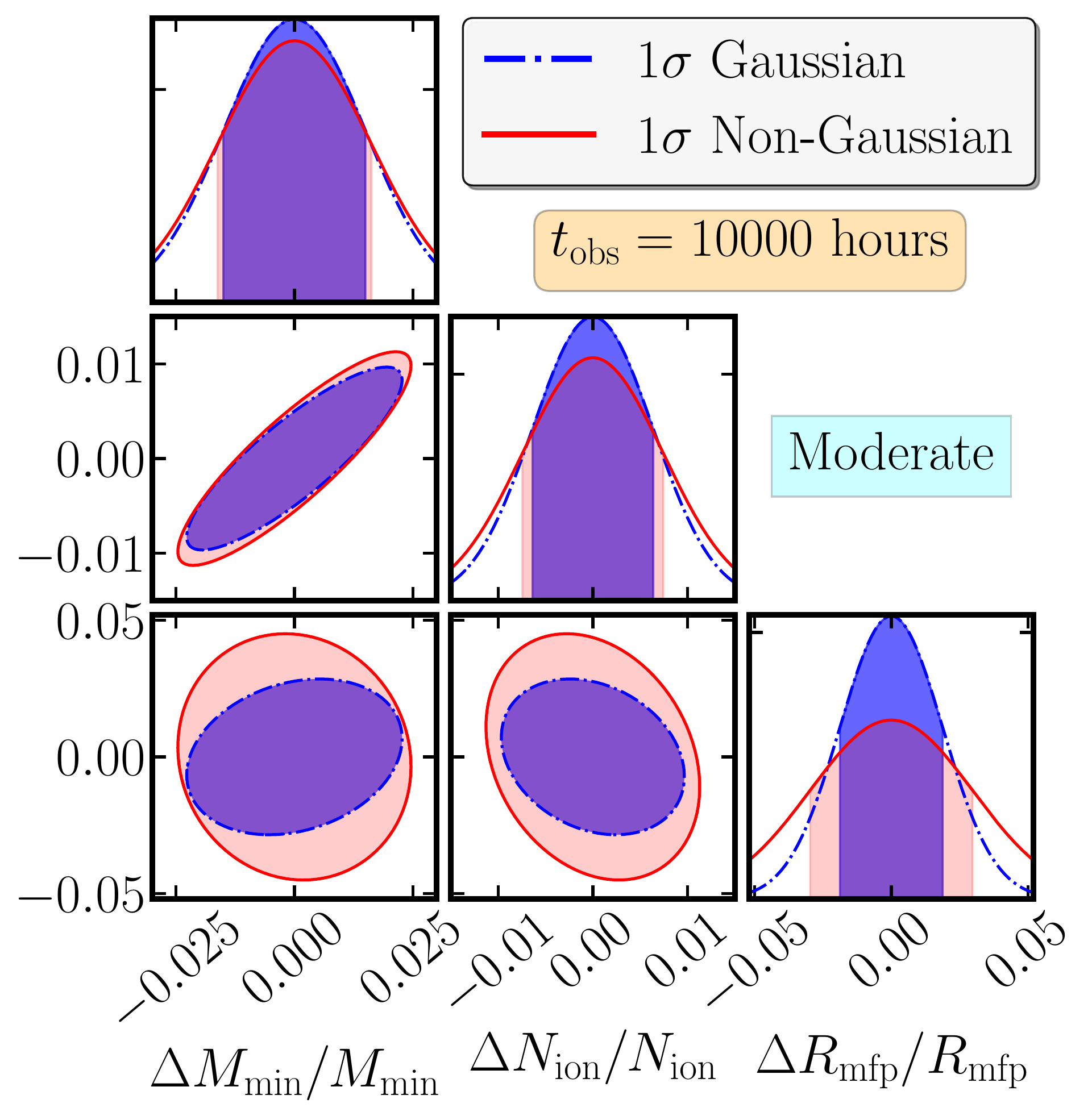}
\includegraphics[scale=0.27]{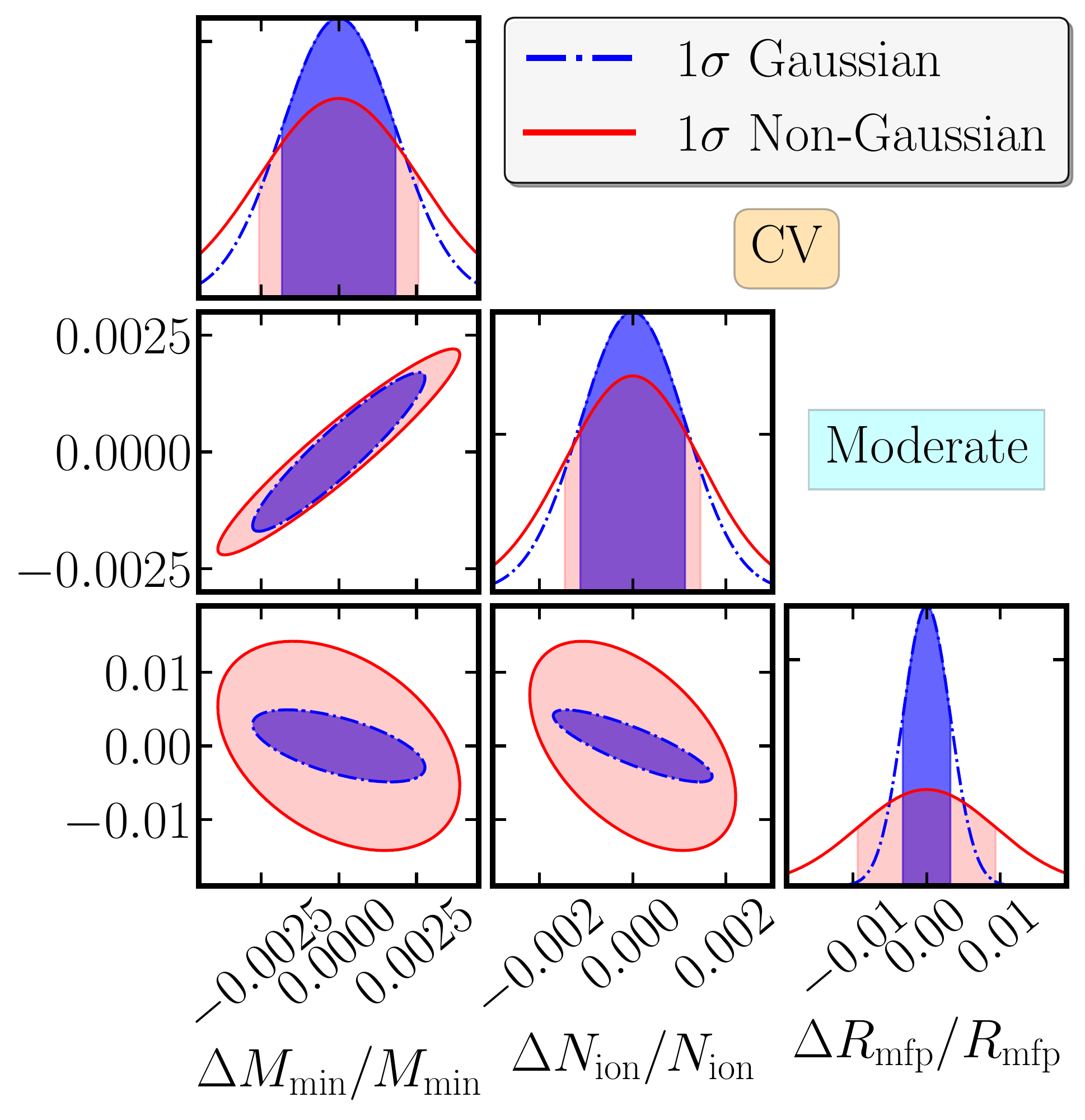}
\caption{The marginalized $1 \sigma$ error ellipses and 1D distribution of fractional errors in parameters for $t_{\rm obs}=1024$ hours (Left), $10000$ hours (Middle) and CV (Right) considering Moderate foreground scenario. This predictions are obtained after combining Fisher matrices for all the six redshift slices.}
\label{fig:elp_obs_mod}
\end{figure*}

Figure \ref{fig:elp_obs_opt} shows the 2D and 1D marginalized errors for the Optimistic scenario. Considering the CV limit (right corner plot) first, we find that the major axes of the non-Gaussian and Gaussian $\Mmin,\Nion$ ellipses are aligned with slopes $31^\circ$ and $30^\circ$ respectively. The non-Gaussian major and minor axes are respectively $3.20$ and $2.63$ times larger than the Gaussian predictions. Considering the $\Mmin,\Rmfp$ ellipses, these shows a mild negative correlation between the corresponding errors with the slopes of the non-Gaussian and Gaussian major axes being $97^\circ$ and $111^\circ$ respectively. Here the ratios of the non-Gaussian and Gaussian major and minor axes are $3.27$ and $4.77$ respectively. Likewise, we see a mild negative correlation between $\DN$ and $\DR$ for both the non-Gaussian and Gaussian predictions where the slopes of the respective major axes are at $96^\circ$ and $105^\circ$. The major and minor axes of the non-Gaussian error ellipse are $4.71$ and $4.88$ times larger than the respective Gaussian predictions. The 1D error predictions are lowest for this case with the non-Gaussian values being $\DM=0.0018$, $\DN=0.0011$ and $\DR=0.0067$, these are respectively $161 \%$, $174 \%$ and $403 \%$ larger than the respective Gaussian predictions. We see that non-Gaussianity has a considerable effect on the error predictions here, the differences being more than $100 \%$ for all the parameters.

The differences between Gaussian and non-Gaussian error predictions are, however, much smaller for both $t_{\rm obs}=1024$ and $10000$ hours. For both these $t_{\rm obs}$, the $\Mmin,\Nion$ error ellipses are inclined at $\sim 20^{\circ}$ to the horizontal, for $t_{\rm obs}=1024$ hours the non-Gaussian major axis is $1.17$ times the Gaussian result while the two minor axes are nearly equal. There is very little difference between the respective axes of the non-Gaussian and Gaussian results for $t_{\rm obs}=10000$ hours. Considering the $\Mmin,\Rmfp$ errors, for $t_{\rm obs}=1024$ hours the non-Gaussian and Gaussian ellipses have slopes of $97^\circ$ and $30^\circ$, whereas the ratios of the respective major and minor axes are $0.95$ and $1.32$. For $t_{\rm obs}=10000$ hours, the corresponding values are $135^\circ$, $55^\circ$, $1.0$ and $0.85$, note that in this case the non-Gaussian error ellipse has a smaller area than the Gaussian one. Considering the $\Nion,\Rmfp$ errors ellipses, for $t_{\rm obs}=1024$ hours the slopes are $97^\circ$ and $82^\circ$ for the non-Gaussian and Gaussian results respectively, whereas the ratio of the corresponding major and minor axes are $1.03$ and $1.10$ respectively. For $t_{\rm obs}=10000$ hours, the corresponding values are $97^\circ$, $92^\circ$, $1.05$ and $0.91$ respectively. Considering the 1D marginalized errors, the non-Gaussian $1 \sigma$ predictions for $t_{\rm obs}=1024$ hours are $\DM = 0.0293$, $\DN=0.0104$ and $\DR=0.0212$ which are respectively $10 \%$, $17 \%$ and $6 \%$ larger than the Gaussian predictions. For $t_{\rm obs}=10000$ hours the non-Gaussian predictions  $\DM = 0.0124$ and $\DN=0.0052$ are respectively $3 \%$ and $17 \% $ larger than the  Gaussian predictions, whereas $\DR=0.0124$ is $18 \%$ smaller than the Gaussian prediction. The marginalized 1D error predictions for all the foreground models and observations times are presented in Table~\ref{tab:3}. To summarize the results for the Optimistic scenario, non-Gaussianity is very important in the CV limit where the error predictions are more than $100 \%$ in excess of the Gaussian ones. The Gaussian system noise dominates the error predictions at  $1024$ hours. We see that even for $t_{\rm obs}=10000$ hours the errors for $\Mmin$ and $\Nion$ are $5$ to $6$ times larger than the CV limit, whereas for $\Rmfp$ the errors are relatively closer to ($1.7$ times) the CV limit. We see that for both $1024$ and $10000$ hours non-Gaussianity can cause differences of at most $\sim 20 \%$ in the 1D error predictions, however this can cause  large difference in the orientation of the 2D error ellipses.

Figure \ref{fig:elp_obs_mod} shows the results for the Moderate scenario. Considering the CV limit, we see that the $\Mmin,\Nion$ error ellipse, both non-Gaussian and Gaussian, have slopes of $\sim 30^\circ$ which matches that of the Optimistic scenario, however the ratio of the respective major and minor axes are  $1.38$ and $1.18$ which are quite a bit smaller than those of the Optimistic scenario. For $\Mmin,\Rmfp$ the non-Gaussian and Gaussian error ellipses have slopes of $96^\circ$ and $112^\circ$ which are very close to the Optimistic scenario, however the ratios of the respective major and minor axes are $1.72$ and $2.73$  which are  quite  smaller than those of the Optimistic scenario. Similarly, for $\Nion,\Rmfp$ we have $94^\circ$ and $107^\circ$ which are very close to the Optimistic scenario, whereas the ratios $2.17$ and $2.79$ are smaller than the Optimistic scenario. The 1D non-Gaussian errors are $\DM=0.0026$, $\DN=0.0015$ and $\DR=0.0094$ which are roughly $\sim 1.5$ times larger than the Optimistic predictions. The non-Gaussian predictions here are respectively $40 \%$, $30 \%$ and $190 \%$ more than their Gaussian predictions, note that in the Optimistic scenario these differences are more than $150 \%$ for $\Mmin$ and $\Nion$, and it is $\sim 400 \%$ for $\Rmfp$. 

\begin{figure*}
\includegraphics[scale=0.27]{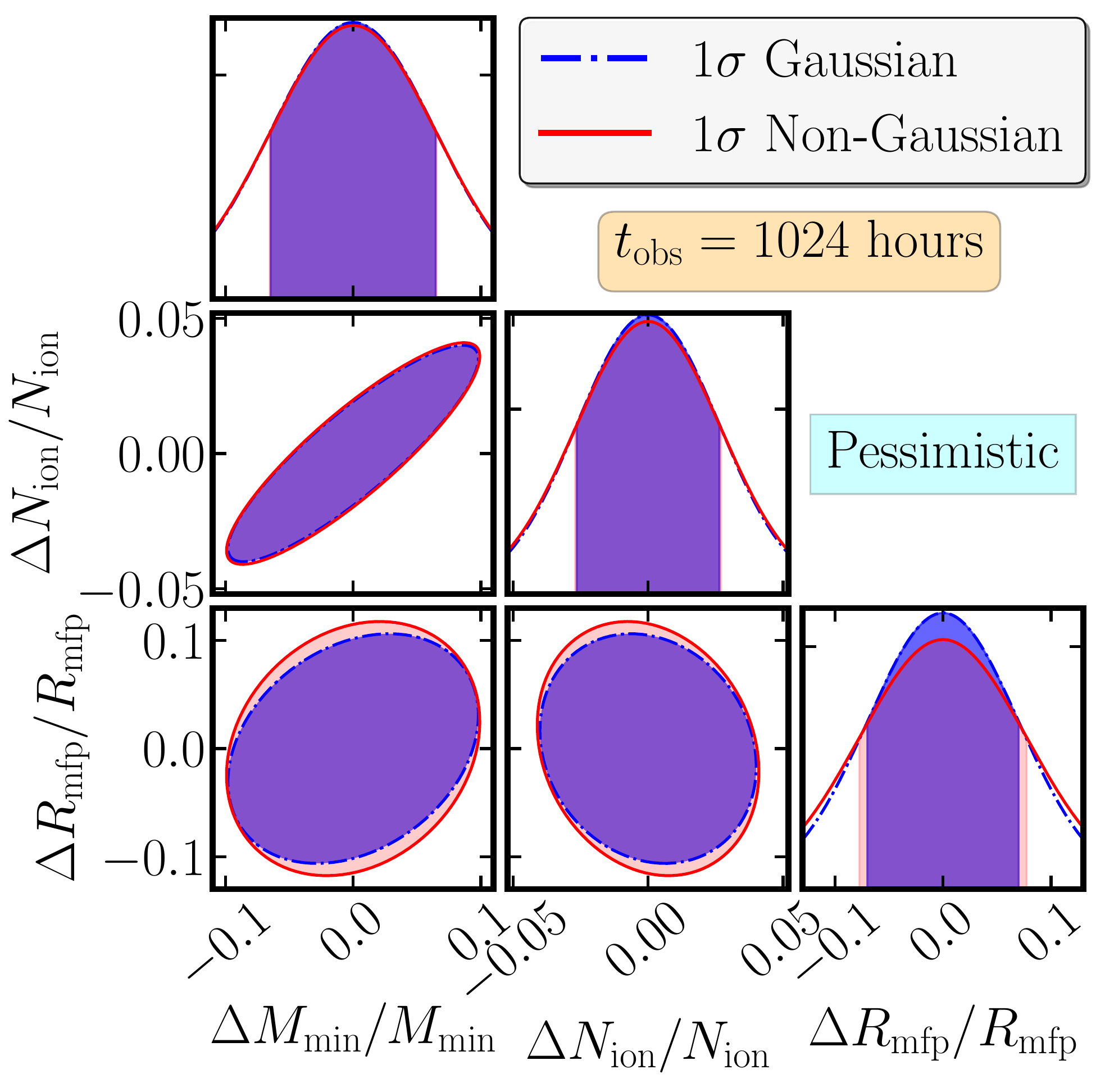}
\includegraphics[scale=0.27]{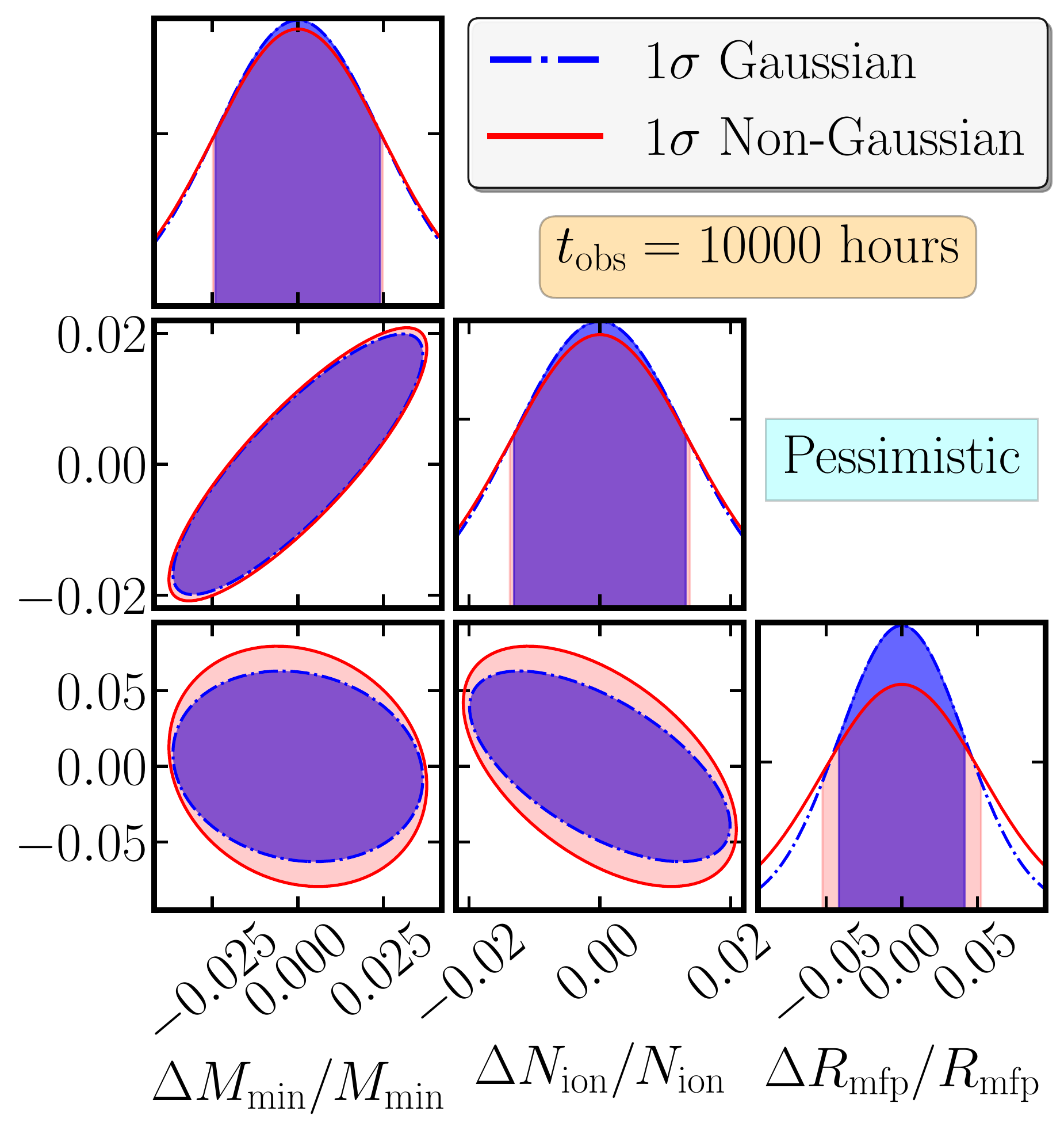}
\includegraphics[scale=0.27]{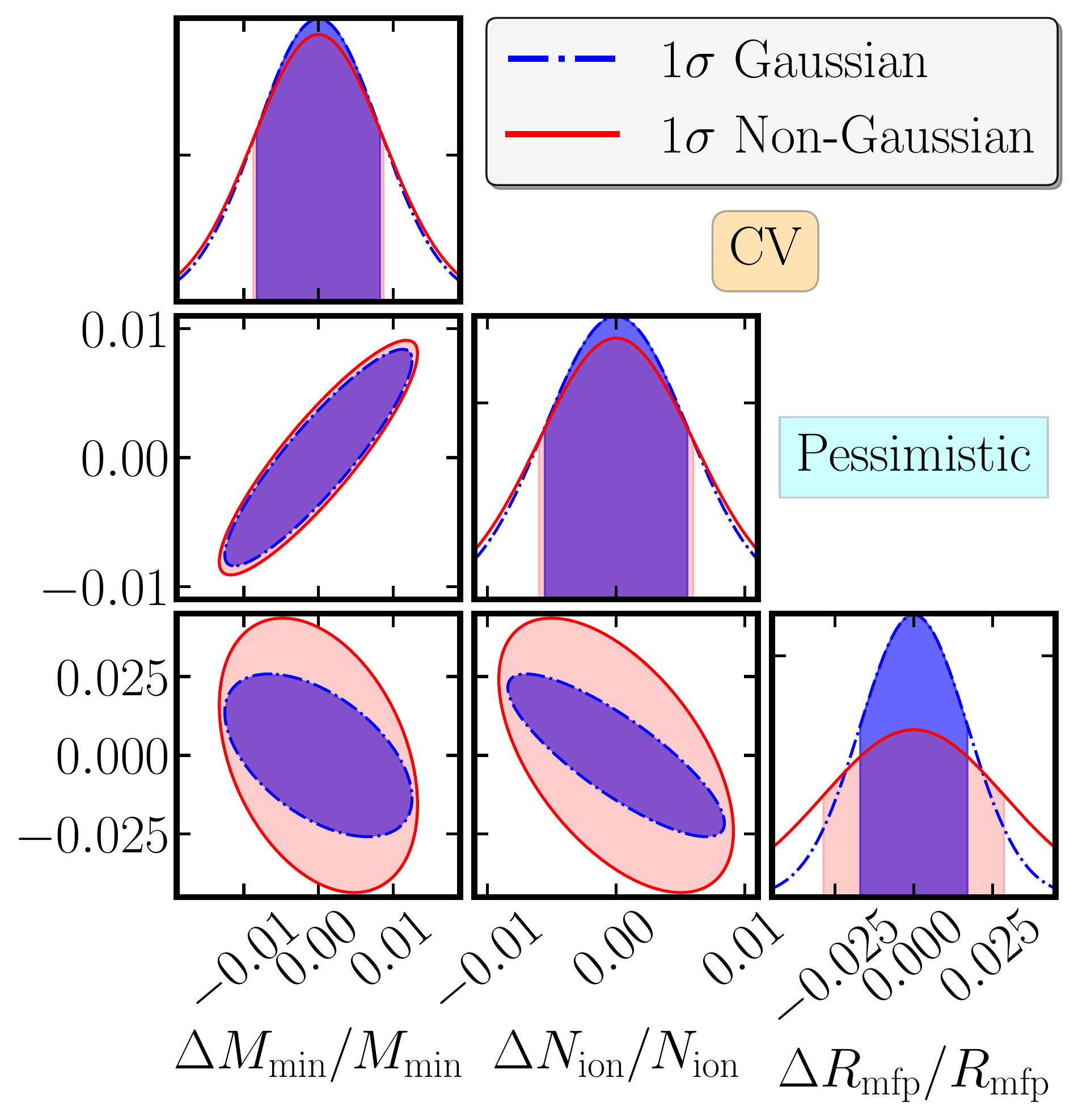}
\caption{The marginalized $1 \sigma$ error ellipses and 1D distribution of fractional errors in parameters for $t_{\rm obs}=1024$ hours (Left), $10000$ hours (Middle) and CV (Right) considering Pessimistic foreground scenario. This predictions are obtained after combining Fisher matrices for all the six redshift slices.}
\label{fig:elp_obs_pes}
\end{figure*}

\begin{table*}
\centering
\begin{tabular}{cc|ccc|ccc|ccc|}
\cline{3-11}
\multicolumn{2}{c|}{} & \multicolumn{3}{c|}{$(\DM)~\times10^{-2}$} & \multicolumn{3}{c|}{$(\DN)~\times10^{-2}$} & \multicolumn{3}{c|}{$(\DR)~\times10^{-2}$} \\ \cline{3-11} 
\multicolumn{2}{c|}{\multirow{-2}{*}{}}  & Non-Gaussian & Gaussian & $\Delta(\%)$ & Non-Gaussian & Gaussian & $\Delta(\%)$ & Non-Gaussian & Gaussian & $\Delta(\%)$ \\ \hline
\multicolumn{1}{|c|}{} & $1024$ & $2.93$ & $2.79$ & $10$ & $1.04$ & $0.96$ & $17$ & $2.12$ & $2.07$ & $6$ \\ \cline{2-11} 
\multicolumn{1}{|c|}{} & $10000$ & $1.24$ & $1.23$ & $3$ & $0.52$ & $0.48$ & $17$ & $1.24$ & $1.38$ & $-18$\\ \cline{2-11} 
\multicolumn{1}{|c|}{\multirow{-3}{*}{Opt}} & CV & $0.18$ & $0.07$ & $161$ & $0.11$ & $0.04$ & $174$ & $0.67$ & $0.13$ & $403$ \\ \hline

\multicolumn{1}{|c|}{} & $1024$ & $3.69$ & $3.55$ & $8$ & $1.36$ & $1.27$ & $14$ & $3.33$ & $2.85$ & $36$ \\ \cline{2-11} 
\multicolumn{1}{|c|}{} & $10000$ & $1.62$ & $1.50$ & $17$ & $0.74$ & $0.64$ & $36$ & $2.97$ & $1.88$ & $150$ \\ \cline{2-11} 
\multicolumn{1}{|c|}{\multirow{-3}{*}{Mod}} & CV & $0.26$ & $0.18$ & $40$ & $0.15$ & $0.11$ & $30$ & $0.94$ & $0.32$ & $190$ \\ \hline

\multicolumn{1}{|c|}{} & $1024$ & $6.54$ & $6.47$ & $2$ & $2.71$ & $2.64$ & $5$ & $7.75$ & $7.00$ & $23$ \\ \cline{2-11} 
\multicolumn{1}{|c|}{} & $10000$ & $2.48$ & $2.41$ & $7$ & $1.38$ & $1.31$ & $10$ & $5.23$ & $4.15$ & $59$ \\ \cline{2-11} 
\multicolumn{1}{|c|}{\multirow{-3}{*}{Pes}} & CV & $0.88$ & $0.83$ & $6$ & $0.60$ & $0.55$ & $8$ & $2.87$ & $1.70$ & $69$ \\ \hline
\end{tabular}
\caption{The $1\sigma$ fractional errors (first two sub-columns) for each inferred parameter considering different foreground models and observation times. Here $\Delta(\%)$ (third sub-column) is the percentage deviation of the non-Gaussian predictions from the Gaussian ones.}
\label{tab:3}
\end{table*}

The error predictions increase considerably when we take the system noise into account. Considering the $\Mmin,\Nion$, for $t_{\rm obs}=1024$ hours the non-Gaussian and Gaussian error ellipses  both have slopes of $\sim 18.5^\circ$ and the respective major and minor axes are nearly equal. The same also holds for $10000$ hours, except that the slope is $\sim 22^{\circ}$. Considering $\Mmin,\Rmfp$, for $1024$ hours the non-Gaussian error ellipse is nearly circular, the ratio to the major and minor axes of the Gaussian error ellipse are $0.94$ and $1.46$ the latter having a slope of $33^\circ$. For $10000$ hours the non-Gaussian and Gaussian ellipses have slopes of $93^\circ$ and $23^\circ$ respectively, whereas the ratios of the respective major and minor axes are $1.15$ and $1.52$. Considering $\Nion,\Rmfp$, for $1024$ hours the non-Gaussian and Gaussian error ellipses have slopes $93^\circ$ and $85^\circ$ respectively, and  the ratios of the respective major and minor axes are $1.07$ and $1.16$. For $10000$ hrs, the respective values are $94^\circ$ and $175^\circ$, $1.58$ and $1.17$. Considering the non-Gaussian 1D errors, for $1024$ hours  we have the $\DM=0.0369$, $\DN=0.0136$ and $\DR=0.0333$ that are respectively $8 \%$, $14 \%$ and $36 \%$ more than the corresponding Gaussian predictions. For $10000$ hours, we have $\DM=0.0162$, $\DN=0.0074$ and $\DR=0.0297$ that are respectively $17 \%$, $36 \%$ and $150 \%$ more than the corresponding Gaussian predictions. We see that in all cases the Moderate scenario error predictions are larger than those for the Optimistic scenario. Here, in the CV limit, the effect of non-Gaussianity is less than that in the Optimistic scenario. However, for both $1024$ and $10000$ hours the effect of non-Gaussianity on the error predictions are larger than those for the Optimistic scenario. This is consistent with the behaviour seen in the values of $\mathcal{R}$ in Table~\ref{tab:2}. The effect is particularly pronounced for $\Rmfp$.

Figure \ref{fig:elp_obs_pes} shows the results for the Pessimistic scenario. Considering the CV limit, we see that for $\Mmin,\Nion$ both the non-Gaussian and Gaussian error ellipses have slopes $\sim 33^\circ$ whilst the ratios of the respective major and the minor axes are $1.13$ and $1.06$. For $\Mmin,\Rmfp$, the slopes of the non-Gaussian and Gaussian error ellipses are $97^\circ$ and $107^\circ$ respectively, while the ratios of the respective major and the minor axes are 
$1.64$ and $1.20$. The corresponding values are $97^\circ$, $106^\circ$, $1.64$ and $1.68$ for $\Nion,\Rmfp$. We see that for all three error ellipses the slopes are similar to those for the Optimistic and Moderate scenarios, however the ratios presented above are smaller than those of the Moderate scenario. Considering the 1D non-Gaussian predictions we have $\DM=0.0088$, $\DN=0.0060$ and $\DR=0.0287$ which are respectively $6 \%$, $8 \%$ and $69 \%$ larger than the corresponding Gaussian predictions. The error predictions increase considerably when we take the system noise into account. Considering the $\Mmin,\Nion$ error ellipse, for $t_{\rm obs}=1024$ hours the non-Gaussian and Gaussian error ellipses are very similar, both have slopes $\approx 20^\circ$ and the respective major and minor axes are nearly equal. The same also holds for $10000$ hours, except that the slope is around $26.5^{\circ}$. Considering $\Mmin,\Rmfp$, for $1024$ hours, the non-Gaussian and Gaussian error ellipses have slopes $65^\circ$ and $53^\circ$ respectively, and the ratios of the respective major and minor axes are $1.05$ and $1.09$. For $10000$ hours, the non-Gaussian and Gaussian ellipses both have slopes of $95^\circ$ and $97^\circ$ respectively, whereas the ratios of the respective major and minor axes are $1.26$ and $1.03$. Considering $\Nion,\Rmfp$, for $1024$ hours, the non-Gaussian and Gaussian error ellipses both have the same slope $94^\circ$, and the ratios of the respective major and minor axes are $1.11$ and $1.02$. For $10000$ hrs, the respective values are $98^\circ$ and $102^\circ$, $1.15$ and $1.25$. Considering the non-Gaussian 1D errors, for $1024$ hours we have the $\DM=0.0654$, $\DN=0.0271$ and $\DR=0.0775$ which are respectively $2 \%$, $5 \%$ and $23 \%$ more than the corresponding Gaussian predictions. For $10000$ hours, we have the $\DM=0.0248$, $\DN=0.0138$ and $\DR=0.0523$ which are respectively $7 \%$, $10 \%$ and $59 \%$ more than the corresponding Gaussian predictions. The error predictions for the Pessimistic scenario are larger than those of the Moderate scenario, the slopes of the $\Mmin,\Rmfp$ and $\Nion,\Rmfp$ 2D error ellipses are also different. For both $1024$ and $10000$ hours, the effect of non-Gaussianity here is $\le 10 \%$  for $\Mmin$ and $\Nion$, however this can be large $(\sim 50 \%)$ for $\Rmfp$.

We see that the $\Mmin,\Nion$  error ellipses for all the foreground scenarios and observations times (including CV) are very similar to the each other and also the corresponding error ellipses for $z\geq 8$ in Figure \ref{fig:elp_cv} where we have separately analysed each redshift without considering the system noise or foregrounds. Note that the $\Mmin,\Nion$ error ellipse for $z=7$ is quite different from those at higher redshifts. We see that most of the information for the $\Mmin,\Nion$ error ellipse comes from the higher redshifts $z \ge 8$ where the non-Gaussian effects are relatively weaker than $z=7$. In contrast, the information regarding $\Rmfp$ only comes from low redshifts ($z=7$) where the non-Gaussian effects are particularly strong. For all the foreground scenarios and observations times, we see that the non-Gaussian effects are particularly important for the 2D error ellipses which involve $\Rmfp$ and also the 1D errors for $\Rmfp$.

We expect the error predictions to fall by a factor of $3.1$ from $1024$ to $10000$ hours of observations in the situation where the covariance matrix $\cov_{ij}$ is system noise dominated. The system noise contribution is relatively stronger at higher $z$ as compared to $z=7$. The error predictions for $\Mmin$ and $\Nion$ are mainly constrained by high $z$ observations, and we expect these to have a relatively larger system noise contribution as compared to $\Rmfp$ which is constrained by observations at $z=7$ only. For the Pessimistic scenario, we see that the 1D error predictions for $\Mmin$ and $\Nion$ fall by factors of $2.8$ and $2.1$ respectively. These are relatively closer to $3.1$ as compared to $\Rmfp$ where the errors only drop by a factor of $1.5$. A similar behaviour is also seen for the other foreground scenarios considered here. 

%======================================================================================%
\section{Summary and conclusions}\label{sec:summ}

The sources and processes which are responsible for ionizing hydrogen in the IGM during EoR, can be modelled through several physically motivated parameters. The PS of the 21-cm radiation from the \HI during the EoR holds the potential to constrain these model parameters. However the statistical errors in the measured 21-cm PS limits the accuracy of the inferred parameter values. Our reionization model has three parameters -- (1) $\Mmin$, the minimum  mass of halos which  can host ionizing sources, (2) $\Nion$, the number of ionizing photons escaping into the IGM per baryon within the halo and (3) $\Rmfp$, the mean free path of the ionizing photons within the IGM. This paper presents error predictions for these three parameters considering future measurements of the 21-cm PS using the upcoming SKA-Low. Several previous works constraining the reionization parameters (e.g. \citealt{Pober_2014,21cmmc,Ewall-Wice,Shimabukuro_2017,Hassan_2017,Kern_2017,Cohen_2018,Binnie_2019,Greig_2019b,Park_2019}) have assumed that the EoR 21-cm signal is a Gaussian random field. However, simulations \citep{Mondal_2015} show that the EoR 21-cm signal is inherently non-Gaussian and the non-Gaussianity increases as the reionization progresses. 

The analysis presented in this paper incorporates the non-Gaussianity of the EoR 21-cm signal. We have used the Fisher matrix to make error predictions for the model parameters of our reionization model. We note that this assumes the errors in the model parameters to have a Gaussian distribution. This assumption is likely to hold given the large number of independent Fourier modes of the 21-cm signal which contribute towards determining the parameter values. However it is important to note that the results presented here, including the ratios of the volumes of the error ellipsoids and the various marginalized error predictions, are all liable to change if this assumption does not hold. The Fisher matrix $F_{\alpha \beta}$ of the three parameters $(\Mmin,\Nion,\Rmfp)$ (equation \ref{eq:fisher}) is related to the partial derivatives of the 21-cm PS with respect to these three parameters and the 21-cm PS error covariance matrix. In this paper we have used simulations to calculate the partial derivatives and the results are presented in Figure \ref{fig:derv}. For the error covariance matrix we have used the results from our recent work (\citetalias{shaw_2019}) where we have considered observations with the upcoming SKA-Low for which we have analysed the effect of non-Gaussianity on the error estimates for the 21-cm PS. The analysis there considers three different foreground scenarios namely (1) Optimistic, (2) Moderate and (3) Pessimistic, which have also been discussed in Section~\ref{sec:fgd} of the present paper. 

The results here are presented in two parts. In the first part (Section~\ref{sub:res_sim}) we ignore all the observational effects arising from the telescope, focusing entirely on the CV which arises from the finite simulation volume and the statistical uncertainties inherent to the signal. We separately consider parameter estimation for the six redshifts $z=13,~11,~10,~9,~8,~7$ where the respective error covariance matrices, which makes no reference to any telescope, were taken from \citet{Mondal_II}. The 21-cm PS is insensitive to $\Rmfp$ at $z > 8$ and the Fisher matrices are 2D whereas these are 3D for $z=8$ and $7$. Non-Gaussianity causes the volume (area) and orientation of the error ellipsoids (ellipses) to differ from the Gaussian predictions. We consider, $\mathcal{R}$, the ratio of the volumes (area) of the non-Gaussian ellipsoid (ellipse) to the Gaussian one to quantify the impact of the non-Gaussianity. We find (Figure \ref{fig:vol}) that $\mathcal{R}$ has modest values in the range $3-4$ during early stages ($z>9$) and it rises gradually to $\sim 6$ for $9 \geq z \geq 8$, beyond which the non-Gaussianity increases abruptly with a very large value ($\mathcal{R}\simeq 70$) at $z=7$. Figure \ref{fig:elp_cv} shows the error predictions for the individual parameters. Considering the $\Mmin,\Nion$ error ellipse we see that for $z = 13$ the non-Gaussian and Gaussian error ellipses both have similar slopes $(\sim 60^{\circ})$. The behaviour is similar for $z \ge 9$, except that the slope decreases to some extent as reionization proceeds and it is $48^{\circ}$ at $z=9$. The behaviour is different at $z=8$ where the slopes are $22^{\circ}$ and $15^{\circ}$ for the non-Gaussian and Gaussian error ellipses respectively. The nature of the error ellipses changes drastically at $z=7$ where the two ellipses both have slopes of $160^{\circ}$. We see that the errors in $\Mmin$ and $\Nion$ are positively correlated at $z \ge 8$ whereas this is negative for $z=7$. The $1 \sigma$ non-Gaussian predictions for $\DM$ and $\DN$ are respectively in the ranges $(0.57-2.53) \times 10^{-2}$, and $ (0.51-4.69) \times 10^{-2}$ (Table \ref{tab:1}), with minimum values at $z=10$ and $8$. The differences between the non-Gaussian and Gaussian error estimates increases as reionization proceeds, except for a dip at $z=8$. Considering $\Rmfp$, this is only weakly constrained for $z=8$. However, at $z=7$ we have $\DR=5.2 \times 10^{-2}$ which is only mildly correlated with the errors in the other two parameters. This non-Gaussian error estimate is $65 \%$ larger than the corresponding Gaussian prediction. 

Predictions for observations with the upcoming SKA-Low are presented in the second part. Note that for each redshift the observational volume is larger than the simulation volume, we have accounted for this in the error covariance matrices. In the second part we have combined the Fisher matrices from all the redshifts to improve the signal-to-noise ratio. The parameters $\Mmin$ and $\Nion$ are mainly constrained by high redshifts where the inherent non-Gaussianity of the 21-cm signal is lower. The system noise is larger for high $z$ observations. In contrast, $\Rmfp$ is mainly constrained by $z=7$ observations, $z=8$ makes a relatively small contribution and the higher $z$ do not contribute at all. Among the redshifts which we have considered, the inherent non-Gaussianity of the 21-cm is largest at $z=7$. We therefore expect the non-Gaussianity to be more important for $\Rmfp$ than to $\Mmin$ and $\Nion$ for both the CV limit and also finite observation times. Considering the Optimistic scenario in the CV limit, the marginalized $\Mmin,\Nion$ non-Gaussian and Gaussian error ellipse both have a slope of $\sim 30^{\circ}$ indicating a positive correlation between the errors in these two parameters (Figure \ref{fig:elp_obs_opt}). The  behaviour is very similar for $t_{\rm obs}=1024$ and $10000$ hours, and also for the other foreground scenarios considered here (Figure \ref{fig:elp_obs_mod} and \ref{fig:elp_obs_pes}). Considering the Optimistic scenario in the CV limit, the $\Mmin,\Rmfp$ and $\Nion,\Rmfp$ non-Gaussian error ellipses both have slopes of $\sim 95^{\circ}$ that indicates negative correlations between the respective errors, the slopes are somewhat larger $(\sim 110^{\circ})$ for the corresponding Gaussian error ellipses. The behaviour is also similar for the other foreground scenarios in the CV limit. This also holds for the $\Nion,\Rmfp$ non-Gaussian error ellipses considering $t_{\rm obs}=1024$ and $10000$ hours, however the slopes of the corresponding Gaussian error ellipses may differ. Considering $\Mmin,\Rmfp$, the slopes of the non-Gaussian and Gaussian error ellipses are different, and these change with $t_{\rm obs}$ and the foreground scenario. Table~\ref{tab:3} summarizes the 1D marginalized errors for all $t_{\rm obs}$ and foreground scenarios. The 1D error predictions are smallest for the Optimistic scenario in the CV limit with the non-Gaussian values being $(\DM,\DN,\DR)=(1.8, 1.1, 6.7) \times 10^{-3}$ which are respectively $161 \%$, $174 \%$ and $403 \%$ larger than the respective Gaussian predictions. The error predictions increase for finite $t_{\rm obs}$ and other foreground scenarios, the effects of non-Gaussianity also come down. The error predictions are largest for $1024$ hours in the Pessimistic scenario where we have $(\DM,\DN,\DR)=(6.54, 2.71, 7.75) \times 10^{-2}$ which are respectively $2 \%$, $5 \%$ and $23 \%$ larger than the respective Gaussian predictions. The error predictions fall and the deviations from the Gaussian predictions increases if $t_{\rm obs}$ is increased. 

In conclusion we note that SKA-Low is predicted to measure the reionization model parameters at $\sim 3 - 8 \%$ accuracy with $1024$ hours of observations in the Pessimistic foreground scenario where the parameters $\Mmin$ and $\Nion$ are not much affected by the non-Gaussianity of the 21-cm signal. However, the errors in $\Rmfp$ and its correlations with the two other parameters are considerably affected by this non-Gaussianity. The accuracy in parameter estimation will increase for longer observations or if the foregrounds contributions can be suppressed further or removed from the data. In this case the effect of non-Gaussianity on the error estimates is expected to increase for all the three parameters. It is therefore important to account for the non-Gaussianity of the EoR 21-cm signal in making realistic predictions for parameter estimation. This will also be important for interpreting future measurements of the 21-cm signal resulting from sensitive upcoming instruments. 

%======================================================================================%
\section*{Acknowledgement}
The authors would like to thank Raghunath Ghara and Srijita Pal for the help related to the SKA-Low baseline distribution in \citetalias{shaw_2019}. AKS would like to thank Anjan Kumar Sarkar, Suman Chatterjee and Debanjan Sarkar for fruitful discussions and comments. RM would like to acknowledge funding form the Science and Technology Facilities Council (grant numbers ST/F002858/1 and ST/I000976/1) and the Southeast Physics Network (SEPNet).

%======================================================================================%
\section*{Data availability}
The data underlying this article will be shared on reasonable request to the corresponding author.

\bibliography{ref}

\label{lastpage}
\bsp	% typesetting comment
\end{document}